\pdfoutput=1

\documentclass[11pt,twoside,a4paper,cmspaper,final,collab]{cms-tdr}
\def\svnVersion{404328}\def\cmsCernNoTag{CERN-EP/2016-320}\def\cmsCernDate{\today}\def\cmsMessage{Published in the Journal of Instrumentation as \href{http://dx.doi.org/10.1088/1748-0221/12/04/P04023}{\doi{10.1088/1748-0221/12/04/P04023.}}}

\begin{document}\cmsNoteHeader{TRK-15-002}

\hyphenation{had-ron-i-za-tion}
\hyphenation{cal-or-i-me-ter}
\hyphenation{de-vices}
\RCS$Revision: 395768 $
\RCS$HeadURL: svn+ssh://svn.cern.ch/reps/tdr2/papers/TRK-15-002/trunk/TRK-15-002.tex $
\RCS$Id: TRK-15-002.tex 395768 2017-03-23 17:57:49Z alverson $
\newlength\cmsFigWidth
\providecommand{\NA}{\ensuremath{\text{---}}\xspace}

\cmsNoteHeader{TRK-15-002}
\title{Mechanical stability of the CMS strip tracker measured with a
  laser alignment system}

\date{\today}

\abstract{
The CMS tracker consists of 206\unit{m$^2$} of  silicon strip sensors assembled
on  carbon fibre composite  structures and is designed for operation in the temperature range
from  $-25$ to $+25^\circ$C.
The mechanical stability of  tracker components during physics
operation was monitored with a few \mum resolution using
 a  dedicated laser alignment system  as well as  particle tracks from cosmic
 rays and hadron-hadron collisions.
During the LHC operational period of 2011--2013 at stable temperatures, the components of the tracker
were observed to experience relative movements of less than  30\mum.
In addition, temperature variations   were found to cause
displacements of tracker structures  of about 2\mum/$^\circ$C, which largely revert to their
initial positions  when the temperature is restored to its original
value.

}

\hypersetup{%
pdfauthor={CMS Collaboration},%
pdftitle={Mechanical stability of the CMS strip tracker  measured with a  laser alignment system},%
pdfsubject={CMS},%
pdfkeywords={CMS, tracker, alignment, lasers}}

\maketitle

\section{Introduction}
The silicon strip tracker of the CMS experiment at the CERN LHC is
designed to provide  precise  and efficient measurements of
 charged particle trajectories in a solenoidal  magnetic field of
 3.8\unit{T} with a transverse momentum  accuracy  of  1--10\%  in the
 range of 1--1000\GeVc in the  central region \cite{trackerjinst}.
It consists of five main subdetectors: the tracker inner barrel with inner disks (TIB and TID),
the tracker outer barrel (TOB), and  the tracker endcaps on positive
and negative sides  (TECP and TECM) \cite{cmsjinst, cmstracker}.
The silicon strip sensors have  pitches varying from 80\mum at the
innermost radial position of 20\cm,  to 205\mum  at the outermost radius
of 116\cm, delivering a single-hit resolution between  10 and 50\mum
 \cite{trackerjinst}.
As a general criterion, the position of the silicon modules has to be
known to much better accuracy than  this intrinsic  resolution.
\par
Silicon sensors exposed to a large radiation fluence require cooling, and the CMS tracker is designed
to operate in a  wide  temperature range from  $-25$ to $+25^\circ$C.
The mechanical stability of the  tracker components  is ensured by the
choice of materials and by an  engineering design that
tolerates the  expected thermal expansion and detector displacements. These
displacements  have to be measured
and accounted for in the form of alignment constants used in the track reconstruction.
\par
The absolute alignment of individual silicon modules is performed with
cosmic ray muons and  tracks from hadron-hadron collisions collected
 during periods of commissioning or collision  data taking
 \cite{cmsalignmentcosmics, cmscommissioning, cmsalignment}.
A significant advance in the track-based alignment came with
the introduction of a global $\chi^2$  algorithm that combines
reconstruction of the track and alignment parameters
\cite{blobel}. This algorithm, implemented in the   \MILLEPEDE  package \cite{mpede},
 was successfully used in various experiments at the LHC, HERA, and
 Tevatron.
The actual accuracy of the track-based alignment depends on the
number of objects requiring alignment and the size of the track
sample.
\par
The movement of the  tracker components over  much shorter  time scales
is  monitored in the CMS experiment with an optical laser alignment system (LAS) \cite{laswittmer}.
Lasers were already  used in the alignment of several silicon-based
tracking detectors, for example, in the ALEPH \cite{alephlas},  ZEUS
\cite{zeuslas}, and  AMS02 \cite{amslas} experiments.
Moreover, the CMS experiment  also uses lasers for linking
the tracker and muon subdetectors together in a common reference frame \cite{cmsmuons}.
There is an alternative method of  optical alignment based on
the RasNiK system that was implemented, for example,  in the CDF \cite{cdfrasnik} and ATLAS
\cite{atlasrasnik} experiments.  The RasNik system uses a conventional
 light source with coded mask, a lens, and a dedicated optical sensor.
Both methods have similar performance, but lasers have some
advantages for operation in the CMS tracker.
First, the infrared laser light  penetrates the silicon sensors, hence
simplifying the alignment system.
Second, the laser light produces a signal  similar to ionizing particles that permits the use
of the same radiation-hard silicon detectors  employed for tracking,
instead of dedicated sensors. Yet another method,  implemented in the
ATLAS experiment, is based on the laser frequency scanning
interferometry \cite{atlasfsi}.
\par
The LAS of the CMS tracker is one of the largest laser-based
alignment systems ever built in high-energy physics.
Forty infrared laser beams  illuminate a subset of 449 silicon modules, and monitor relative displacements of the TIB,
TOB, and TEC subdetectors over  a time interval of a few minutes
with a stability of a few \mum \cite{laswittmer}.
Alignment with particle tracks and laser beams  are complementary
techniques and together they ensure the high quality of  track
reconstruction.
While the track-based alignment is used to reconstruct the alignment
constants of individual modules, the LAS  identifies short-term  displacements of large
structures in order to exclude such
periods from the offline analysis of the experimental data.
\par
In this paper we describe the   mechanical structure of the tracker and the  LAS  in detail.
We review the alignment procedure of using laser beams and particle tracks.
The  measurements of the mechanical stability of the tracker components
during the LHC data taking  period in 2011--2013,  as well as during the LHC long
shutdown period spanning 2013--2014, are  presented and discussed.

\begin{figure*}[thb]
  \centering
    \includegraphics[width=0.95\textwidth]{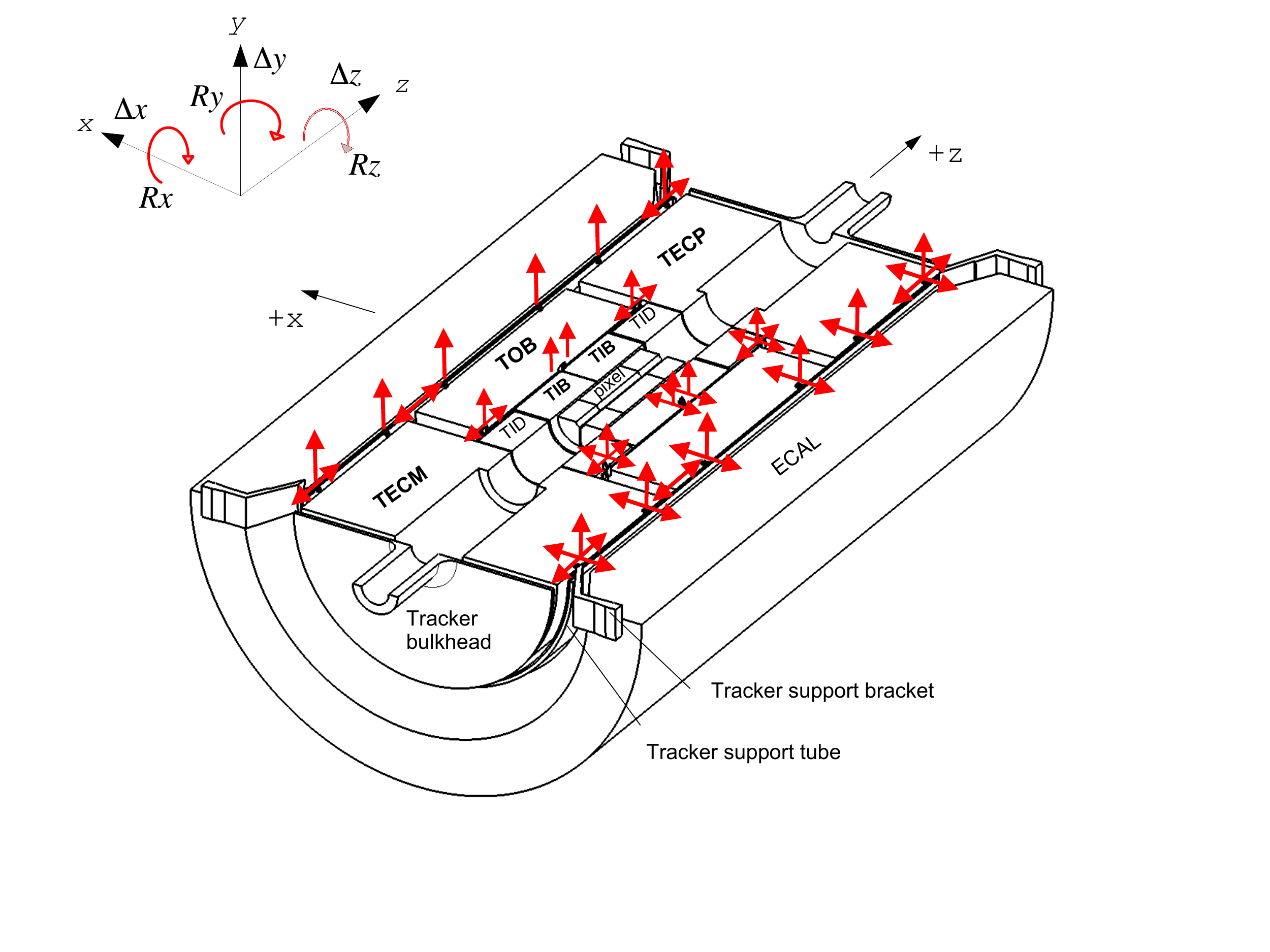}
    \caption{Mechanical layout  and mounting of the tracker
      subdetectors (bottom half is shown).
The TIB+TID are mounted inside the TOB, while
the TOB, TECP, and the TECM are mounted inside the TST.
The red arrows indicate the connection points and their kinematic constraints.}
      \label{fig:trackerlayout}
 \end{figure*}

\section{Mechanical design of the  CMS tracker}
The silicon strip tracker of the CMS detector  is composed of  15\,148 silicon strip detector modules
with a total area of about 206\unit{m$^2$} and is described in Refs.
\cite{cmsjinst, cmstracker}.
Below we discuss in more detail the components of the tracker
 that are relevant to the mechanical stability of the detector.
The mechanical concept of the  tracker is  sketched in Figure \ref{fig:trackerlayout}.
The CMS coordinate system has its origin at the centre of the
detector with the $z$-coordinate  along the LHC beam pipe, in the direction of the counterclockwise proton beam,  and the
horizontal $x$- and the vertical  $y$-coordinates  perpendicular to
the beam (in the cylindrical system $r$ is the radial distance and  $\varphi$ is the azimuth).
The inner radii from 4.4 up to 15\cm are occupied by the silicon pixel
detector,  which is operated independently of the strip tracker.
The silicon strip  modules are mounted around  the beam pipe at  radii
from  20\cm to 116\cm  inside a cylinder of 2.4\unit{m} in
diameter and 5.6\unit{m} in length.   The TIB  extends in $z$ to  $\pm$70\cm and in radius to 55\cm.
It is composed of two half-length barrels with four detector layers,
supplemented by three TID disks  at each end. The TID disks are equipped
with wedge-shaped silicon detectors with radial strips.
The TOB surrounds the TIB+TID. It has an outer radius of 116\cm,
ranges in $\abs{z}$ up to  118\cm, and consists of six barrel layers.
In the barrel part of the tracker, the
detector strips are oriented along the $z$-direction, except for the double-sided
stereo modules in the first two layers of the TIB and TOB, where they are rotated at an angle of 100\unit{mrad},
providing reconstruction of the $z$-coordinate.
The TECP and TECM cover the region $124 < \abs{z} <282$\cm and $22.5 < r < 113.5$\cm.
Each TEC is composed of nine disks, carrying up to seven rings of  wedge-shaped silicon detectors
with radial strips, similar to the TID. Rings 1, 2, and 5 are also equipped
with  stereo modules for reconstruction of the $r$-coordinate.
\par
Each module of the silicon strip detector  has one or  two silicon sensors
that are  glued on carbon fibre (CF) frames together with a ceramic readout hybrid,
 with a mounting precision of 10\mum.
Overall, there are 27 different module designs optimized for
different positions in the tracker.
The detector  modules are mounted on  substructures that are, in turn,  mounted on the tracker subdetectors.
\par
The TIB is split into two halves for the negative and positive $z$-coordinates allowing easy insertion into the TOB.
The  TIB substructures consist of 16 CF half-cylinders, or
 shells. The mounting accuracy  of detector modules on
the shells is about  20\mum in the shell plane.
The modules are assembled in rows that overlap like roof tiles for better coverage and
compensation for the Lorentz angle \cite{cmstracker}.
An aluminium cooling tube, with 0.3\mm wall thickness and 4$\times$1.5\unit{mm$^2$}
rectangular profile  is glued to the mounts of the detector modules.
Each row has three  modules on one cooling loop and
each cooling pipe is connected at the edges of the  shells to the circular
collector pipe that gives extra rigidity to the whole TIB mechanical
structure.  The overall positional accuracy of the assembly of all
shells is about 500\mum.
\par
The TOB main structure  consists of six cylindrical layers supported by
four disks, two at the ends and two in the middle of the TOB
structure. The disks are made of 2\mm thick CF laminate and are connected by cylinders at the inner and outer diameters. The cylinders  are produced from  0.4\mm CF skins
glued onto two sides of a 20\mm thick  aramid-fibre  honeycomb core.
The  detector modules are mounted on 688 substructures called rods.
The rods are inserted into  openings on the disks, such that each rod is supported by two  disks.
The  accuracy of mounting the rods is about 140\mum in $r$-$\varphi$
and 500\mum  in $z$.
Each rod has 6 or 12 (for rods with double-sided modules) silicon modules mounted in a row.
A 2\mm diameter copper-nickel  cooling pipe is attached to the
CF  frame of the rod.
Each module is mounted on the rod with an accuracy of 30\mum by two
precision  inserts   connected  to the cooling pipes, and two springs.
\par
Each TEC side consists of nine disks  with 16 wedge-shaped substructures on each disk, called petals.
Overall there are 144 petals with different  layouts, depending on the disk location.
The petals are made of CF skins with a honeycomb structure inside.
The wedge-shaped  detector  modules are mounted on the petals   with an accuracy of 20\mum
using four aluminium  inserts that are  connected to the cooling pipe.
A titanium cooling pipe of about 7\unit{m}  in length, 3.9\mm in diameter,
with 0.25\mm wall thickness  is integrated into the petal honeycomb structure and  is bent to
connect all heat sink inserts. The petals are  mounted on the CF disks
with a  precision of 70\mum. All  nine disks of each TEC are connected together with  eight
CF bars forming a rigid structure. These bars are also used to
hold service cables and cooling pipes.
The  overall accuracy of the disks assembly in the TEC
subdetector is about 150\mum in  all coordinates.
\par
The main support structure for all tracker subdetectors is the tracker
support tube (TST).
The TST is a cylinder 5.4\unit{m} in length and 2.4\unit{m} in diameter made of CF composite.
The  wall of the TST is made of a 30\mm thick  sandwich structure
with   2\mm CF skins on both sides, and a 26\mm thick aluminium honeycomb core.
The TOB, TECP,  and the TECM are  supported inside the TST while the
TIB and TID are supported by the TOB.
The total weight of all subdetectors inside the TST  is about 2200\unit{kg}, which is distributed on two  longitudinal  rails
connected  to the TST with glue and metallic inserts.
The TST itself is supported inside the CMS calorimeters   by four  brackets at each end.
According to calculations the  maximum  deformation of the TST when
supporting the assembled tracker is about 0.6\mm.
The mounting  accuracy of  different subdetectors inside the TST is in the  range of  1\mm, but
the exact position of all subdetectors  was  measured with an accuracy
of 50\mum  in an optical survey  conducted at the beginning of the
detector operation \cite{cmscommissioning}.
A possible movement that is beyond the assembly accuracy is
considered as a major displacement.
\par
The detector modules, substructures, and subdetectors are joined
together using  the so-called kinematic connections that
constrain the movement in some directions, where
the constraints are ensured by the static friction in tension screws.
The engineering  designs of these connections in the various mechanical structures are different, but
the range in all joints suffices to accommodate  the expected thermal movement.
The  fixation points and allowable movements for the subdetectors  are
indicated in Figure \ref{fig:trackerlayout}.
In the vertical direction, each subdetector is constrained only by its own weight.
The movement in the $x$-direction is constrained  only on one side of
the TST. The fixations in the  $z$-coordinate are governed by the  assembly procedure.
During the assembly, the TOB was first inserted into the TST and fixed in
$z$ on one side. Then the TIB and TID  halves were inserted into the
TOB from each end and fixed against each other in the  centre. The
TECP and TECM were mounted last, and constrained in $z$ at the internal ends.
\par
Operation of  silicon modules exposed to a large radiation fluence  requires  cooling \cite{cmstracker}.
The total dissipated power of the readout electronics  with a fully powered tracker is about 45\unit{kW}.
After  irradiation the leakage current of the silicon sensors  contributes another 10\unit{kW}.
The heat inside the tracker is evacuated by a monophase liquid-cooling
system that uses a fluorocarbon (C$_{6}$F$_{14}$)  coolant.
Two cooling plants, each with  40\unit{kW} capacity,  are used for this purpose. Each plant  is
connected to 90 cooling loops distributed among the different
substructures in one half of the tracker.
\par
In the first physics run during  2010--2013 (Run 1),  the operating
temperature of the cooling plant   was set to $\mathrm{T} = +4^\circ$C.  For
Run 2 (2015 onwards),   the nominal operating temperature  was
decreased to  $-15^\circ$C,  in order to allow long term operation  with
increased  fluence  caused by increased energy of collisions, as well as
instantaneous luminosity \cite{cmstracker}.
The operation at low-temperatures  requires a low dew point inside the
tracker. The whole tracker volume is separated from the TST inner wall by a thermal
screen, apart from the points at which the subdetectors are attached to the rails that
remain at ambient temperature. The thermal screen
has cooling elements inside and heating elements outside the tracker
volume, thus acting as a thermal barrier that prevents condensation.
The inner  tracker  volume of about 25\unit{m$^3$} is constantly flushed  with dry air or nitrogen
at a rate of about 20\unit{m$^3$/hour}, such that
the dew point in the CMS cavern of  about  $+10^\circ$C is
reduced to  below $-40^\circ$C inside the tracker volume.
All service cables and cooling pipes leave the subsystem at the tracker
bulkheads, which are also isolated by the thermal screen and flushed with
dry gas at a higher rate of about 150\unit{m$^3$/hour}.
\par
\begin{figure*}[thb]
  \centering
    \includegraphics[width=0.85\textwidth]{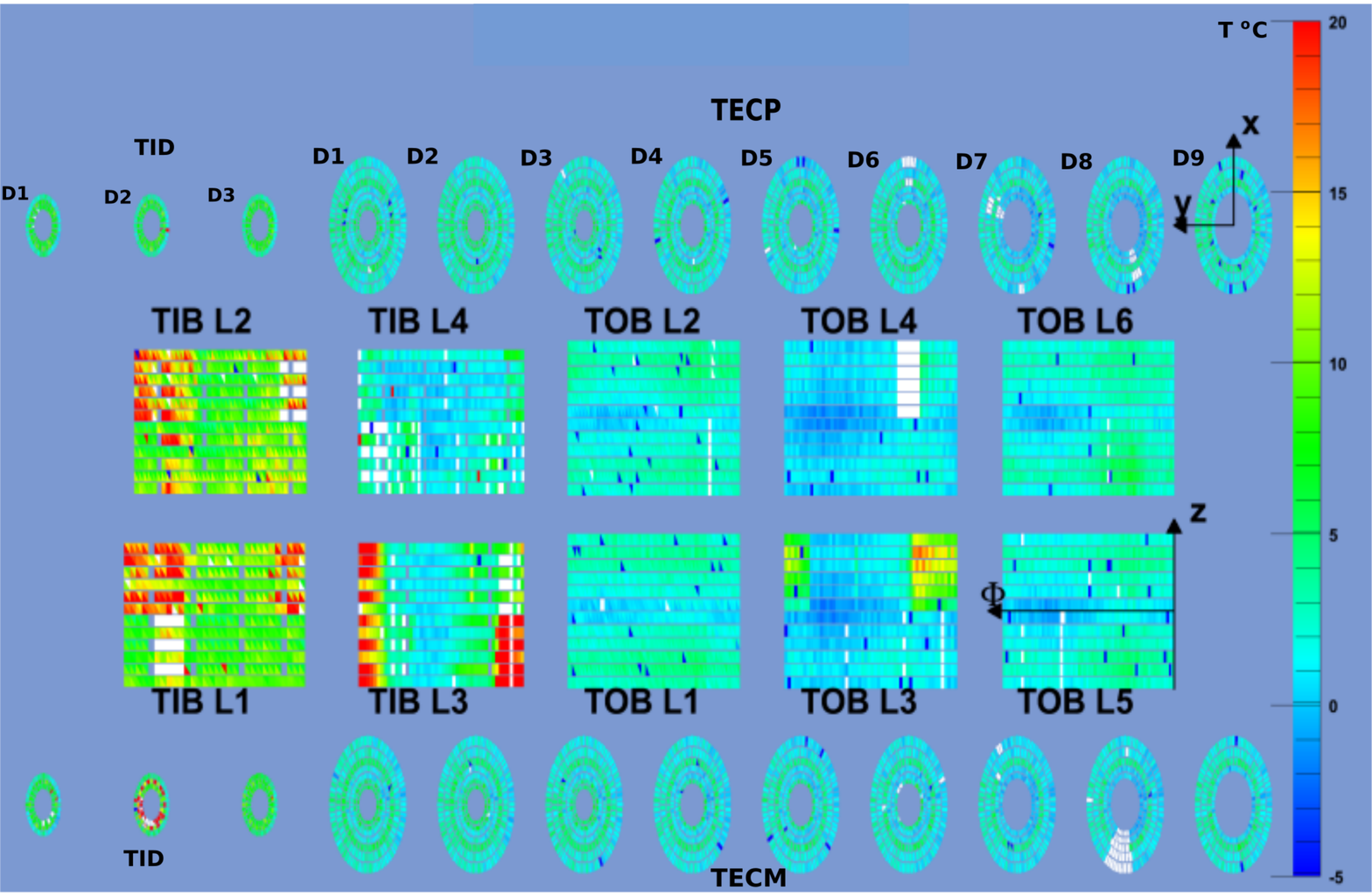}
    \caption{Example of the temperature distribution, shown as a
      colour palette ($^\circ$C),
               measured on silicon sensors in  the TIB (L1--L4),
      TOB layers (L1--L6), and the TEC (D1--D9),  TID (D1--D3)  disks,   with the
      cooling plant operating at $\mathrm{T} = -5^\circ$C. The white spots
      correspond to  nonoperational detectors, and  red spots are the
      closed cooling loops and bad cooling contacts.}
      \label{fig:tempdist}
 \end{figure*}

The  temperature of the  different mechanical structures inside
the tracker depends on the distribution of heat sources and
heat sinks.
The temperature and humidity inside the tracker are monitored by  dedicated
sensors mounted  directly on readout hybrids,  silicon sensors, and
mechanical structures,  distributed throughout the detector volume.
The nonuniformity  of heat dissipation and heat removal results
 in significant  temperature variations inside the tracker even in thermal equilibrium.
Large temperature gradients are  observed near the readout hybrids,
the cooling tubes and the mounting connections.
Figure \ref{fig:tempdist} shows the temperature  measured on
silicon sensors in different subdetectors running with a cooling plant
at operating temperature of  $-5^\circ$C.
The white areas represent non-operational detectors, which comprise
about $2.5\%$ of the total area.
The red (hot) spots are five cooling loops (three in the TIB, and one in the
TOB and the TID) that are closed because of leaks and bad cooling
contacts  (layers L1, and  L2 in the TIB).
\par
A local change of temperature naturally causes an increase in the nonuniformity.
For example, the powering of the module readout electronics rapidly increases the  local
temperature  by  about 15$^\circ$C and
it takes about one hour to stabilize the temperature in the tracker volume.
Cooling down from the ambient temperature of
about $+15^\circ$C to  $+4^\circ$C takes about 3 hours before
stabilization of  temperature.

\section{The laser alignment system}
\par
The initial  purpose of the LAS was to measure relative positions of
the tracker subdetectors  with an accuracy of
about 10\mum and the absolute position with an accuracy of 100\mum.
The large temperature variations expected in the tracker
determined the design concept and  components for the LAS;
the components had to be light, radiation hard,
operational in a high magnetic field, and capable of sustaining large temperature variations.
\par
\begin{figure*}[thb]
  \centering
    \includegraphics[width=0.9\textwidth]{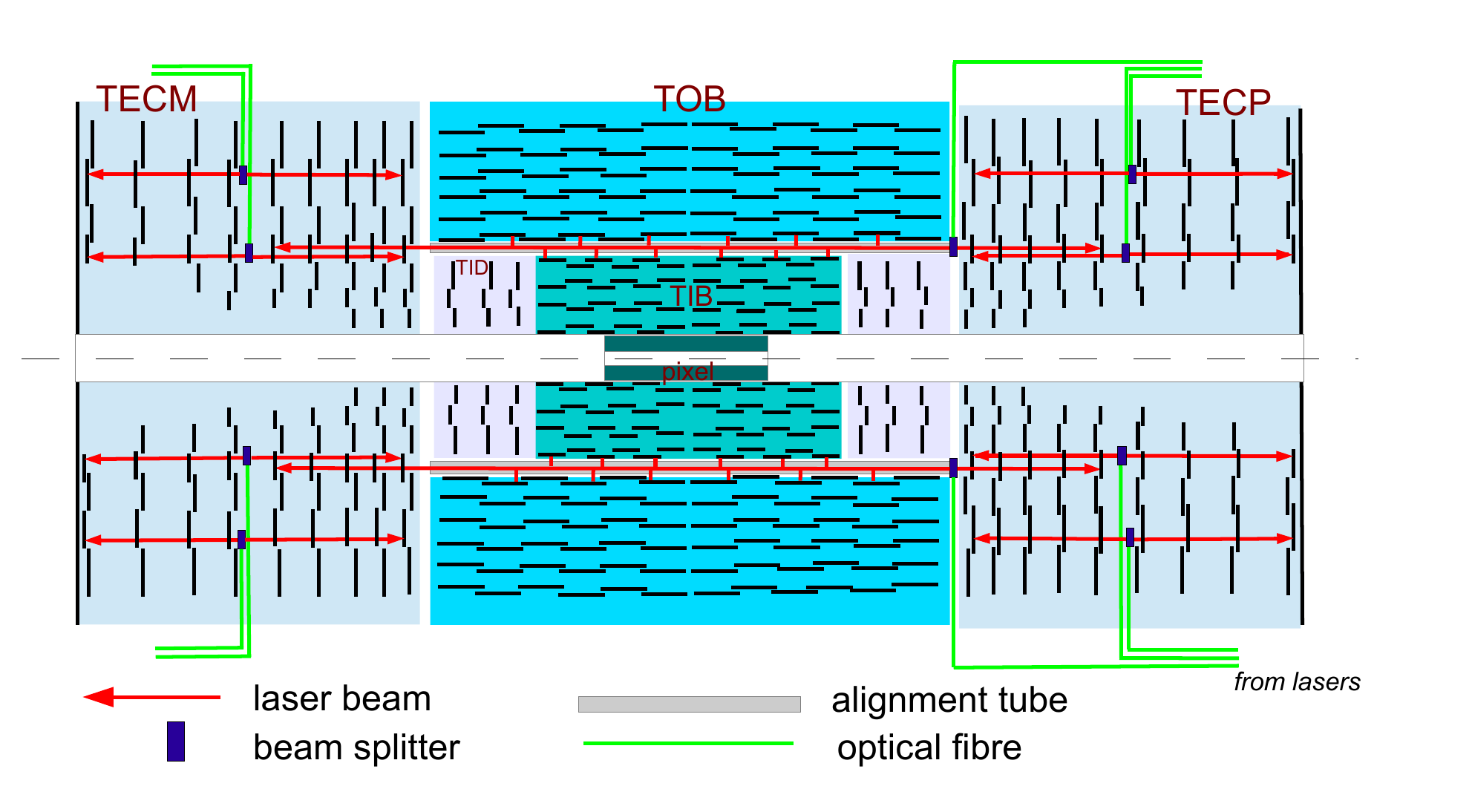}
    \caption{Distribution of the laser beams in the CMS tracker. The
      eight laser beams inside the alignment tubes  are used for the
      global alignment of  TOB, TIB, and TEC subdetectors. The 32 laser
      beams in the TECs are used for the internal alignment of TEC disks.}
      \label{fig:laslayout}
 \end{figure*}

The LAS has 40 infrared laser beams that illuminate the
silicon strip modules in the outer layer of the TIB, inner layer
of the TOB, and in rings 4 and 6 of the TECs, as shown in Figure \ref{fig:laslayout}.
The LAS uses the same detector modules that are  used for particle
detection. Laser pulses are triggered during the  3\mus orbit  gap
corresponding to 119 missing bunches in the LHC beam structure that
has an orbit time of 89\mus, thus not interfering with collisions \cite{cmsjinst}.
The laser beams  are split into two sets.
Eight beams are used for global alignment of the TIB, TOB, and  the TECs
relative to each other.
The other 32  beams are used to internally align the disks in the TECP and TECM subdetectors.
The pixel and the TID subdetectors are not
included in the LAS monitoring.
Since the tracker has approximate axial symmetry, the beams are
distributed rather uniformly in the $\varphi$-direction, with
the exact position  defined by the mechanical layout.
Each laser beam used for global alignment  illuminates six modules in
the TIB,  six modules in the TOB, and  five modules in each of the  TECP
and TECM subdetectors.  Each laser beam for the internal TEC alignment traverses
nine modules in each TEC subdetector.
\begin{figure*}[!thb]
   \centering
  \includegraphics[width=0.8\textwidth]{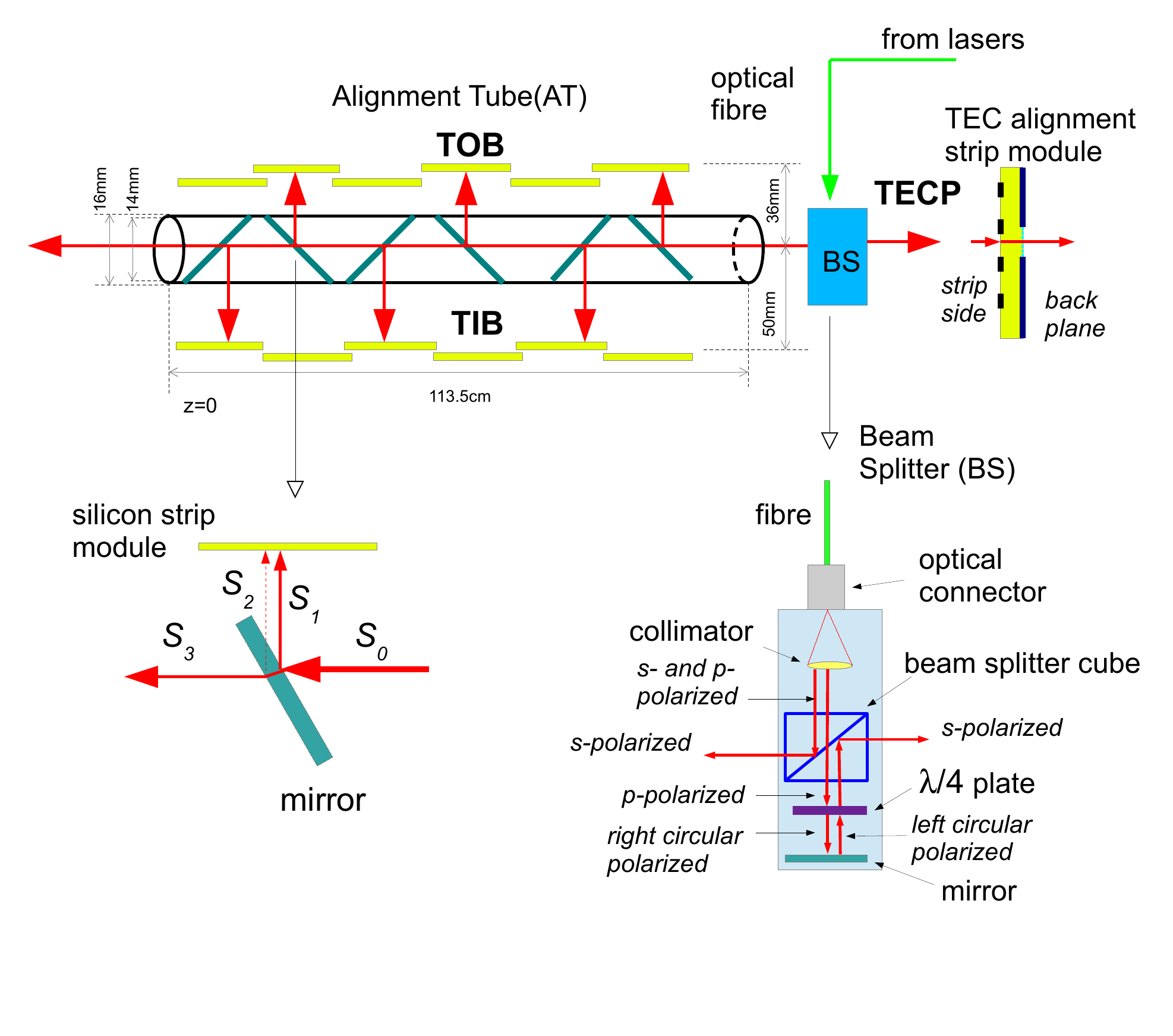}
     \caption{The LAS components: alignment tube, mirror, and beam
       splitter.}
       \label{fig:lasdetails}
  \end{figure*}
\par
The LAS components include laser diodes, depolarizers, optical fibres,
beam splitters, alignment tubes, mirrors,  and specially treated silicon sensors.
The laser  diodes  QFLD-1060-50S   produced by
QPhotonics have a wavelength  of $\lambda = 1075\pm$3.5\unit{nm}
and a maximum optical power of 50\unit{mW}. The attenuation
length of this laser light in silicon  is 10\cm  at $0^\circ$C and  decreases  by
$1\%$/$^\circ$C with increasing temperature.  The light output of each
laser  is regulated by an operating current in the range of 0--240\unit{mA}
and is optimized  as discussed below. The lasers  operate in pulsed mode with a pulse width of 50\unit{ns}.
The spectral bandwidth of the lasers is $\Delta \lambda = 2.4\pm0.9$\unit{nm}, which defines the
coherence length $\lambda^2/\Delta \lambda\approx480\mum$.
A coherence length  larger than 2d$\cdot$n$_\mathrm{Si}$ (where d is the silicon
thickness of 320--500\mum and n$_\mathrm{Si}\approx3.5$ is the silicon refractive index)
would result in  interference of the laser light
reflected on the front- and  back-side of the sensor, hence degrading
the laser beam profile.
Because of the harsh radiation conditions in the CMS detector cavern,
the laser diodes are located in the outer underground service area. The
light is distributed to different subdetectors
 via  special 0.125\mm diameter Corning monomode optical fibres.
\par
The light from the lasers is  directed towards different subdetectors
using beam splitters (BS) that divide the
light into back-to-back beams, as shown in Figure
\ref{fig:laslayout}. For the global alignment the eight beam splitters
for the eight laser beams are mounted between the TOB and TECP. For
the internal TEC alignment,  32 BS are located on disk
6 in both  TECP and TECM.
The principle of operation of the beam splitter is based on
polarization using  a $\lambda$-plate, as shown in Figure \ref{fig:lasdetails}.
The incoming beam, consisting of both $s$- (perpendicular to the
plane of incidence)  and $p$- (parallel) polarizations,  is collimated
onto a $45^\circ$ inclined surface with a special coating, from which the $s$-polarized part of the
laser light is completely reflected. The $p$-fraction continues,
traversing a $\lambda/4$-plate and converting into right circular
polarization. The light is reflected by a mirror after the plate and
changes polarization to  left circular. After a second traversal of
the $\lambda/4$-plate, the left circularly polarized light becomes
 $s$-polarized and is completely reflected onto the other side of the 45$^\circ$ inclined surface. At the end, there  are two
parallel back-to-back beams with $s$-polarization in the $z$-direction. The important
characteristic of the splitter is the variation of collinearity
as a function of the beam spot position. For  all BS, the
 collinearity is measured to be less  than  50\unit{$\mu$rad} for the  $-25$ to  $+25^\circ$C  temperature range.
Since the  laser light is polarized, splitting based on
a $\lambda$-plate  requires  depolarization.  The depolarizers
(produced by  Phoenix Photonics) are located in the
service area just after the laser diodes and  before the optical fibres.
\par
Dedicated  alignment tubes (AT) between the TIB and TOB  are used to hold
the BS  and semitransparent mirrors that reflect light
towards TOB and TIB detector modules, as shown in Figure \ref{fig:lasdetails}.  The eight laser beams between TOB and TIB
pass through the AT and continue to the TECM disks.
The AT are made from  16\mm diameter aluminium and
are integrated into the TOB support wheels.
The mirrors  mounted inside the AT are glass plates that reflect  about 5\% of
the light intensity perpendicular to the beam ($S_1$). The  antireflective
coating on the back side  of the mirror and the $s$-polarization of
the laser light after the beam splitter prevent the second reflection
($S_2$).
The mounting accuracy of each AT is about 100\mum, but
with temperature variations the aluminium  can expand by about
0.5\unit{mm}/m/$20^{\circ}$C.  Although this expansion is mostly along the
$z$-direction, the movement can affect the orientation of the BS and therefore the direction of the laser beams. 
Such variations are  taken into account in the LAS reconstruction
procedure, as discussed below.
\par
Overall  the laser beams hit  449 silicon sensors, with a strip
pitch varying from 120\mum in the  TIB to 156\mum in the TEC detector modules.
The 48 TIB and 48 TOB sensors that are used by the LAS are standard
ones, and are illuminated on the strip side. On the other hand, 353 TEC sensors  had to be  modified to allow  the passage of laser light.
For the standard sensors, the backplane is covered with aluminium
coating and is therefore not transparent to the laser light.
For the TEC modules this coating was removed  in a 10\mm diameter
circular area of the anticipated laser spot position. In addition,  an  antireflective coating was applied in this area in
order to improve the transparency. An attempt also to coat  the strip
sides resulted in changes of the silicon sensor electrical properties and
was therefore abandoned.
\par
Since the detector modules  illuminated by lasers are also used for
particle tracking, their readout electronics is exactly the same as
for other  silicon strip modules in the tracker \cite{cmstracker}.
The signal from the silicon strips  is processed in the analogue
pipeline readout chip (APV25) \cite{cmsjinst},
and transferred to the data acquisition (DAQ) by  optical fibres.
The analogue signal from the APV25 chip  is digitized by the
analogue-to-digital converters (ADC)  located in the CMS underground
service cavern, and is processed further similarly to  physics data.
The LAS-specific electronics  include
a trigger board that is  synchronized with the  CMS  trigger system
and the 40 laser drivers.
\par
The trigger delay for each laser driver is tuned individually to
ensure that the laser signals arrive  at the detector module properly  synchronized with the CMS readout sequence.
The laser intensity is also optimized individually  to account for
losses in the optical components and attenuation in the TEC silicon
sensors. The  amplitude and time settings for each laser driver  are defined in
a special calibration run, during which the laser  intensity and delays are scanned in small steps.
There are  five settings available for each laser, to be shared
between up to 22 detector modules illuminated by the same laser beam.
This results in some variations of the laser signal amplitude in different detectors.
\par
One regular LAS  acquisition step  consists of 2000 triggers.
The first 1000 triggers are optimized for the global alignment, whilst
the other 1000 triggers are used for the internal  alignment of both
TECs. The lasers are triggered with five different settings delivering
200 suitable laser shots for each illuminated module.
The signal-to-noise ratio for the 200 accumulated pulses is above 20, which is similar to the signal from particle tracks.
The lasers are triggered in the orbit gap of every hundredth LHC beam
cycle, corresponding to a rate of 100\unit{Hz}  and resulting in about 20\unit{s}
per  acquisition step. 
During normal data taking the acquisition interval was set to 5\unit{minutes} to
achieve a good compromise between the time resolution of the LAS alignment and the stored data volume.
Since the LAS electronics is deeply integrated into the CMS data acquisition,
it works only when the tracker and the DAQ
are  operational and configured for a global physics run. Intervals
between the runs, periods of testing, and technical stops are not
covered by the LAS measurements.

\section{Tracker alignment}
The general  tracker alignment procedure reflects the mechanical structure of the detector.
The largest alignable objects are the tracker subdetectors, and
the smallest ones  are the silicon sensors.
Each alignable object is considered as an independent and mechanically rigid  body that can move
and rotate in  six degrees of freedom: three offsets ($\Delta x,
\Delta y, \Delta z$)  and three rotations ($Rx,Ry,Rz$) around the axes,
as shown in Figure~\ref{fig:trackerlayout}.
\par
The alignment procedures used with particle tracks and  the LAS data differ somewhat.
In  the LAS, the assembly accuracy  and the mechanical stability of
the optical components are about 100\mum,  limiting the accuracy of
absolute alignment  to  about 50\mum \cite{laswittmer}.
Relative displacements  with respect to a reference
position can be monitored using LAS data with a much better precision
of a few \mum. However, the limited number of laser beams only
allows the reconstruction of the relative
displacement of large structures, such as the TOB, TIB, and the
TECs,  using some assumptions discussed below.
\par
The alignment with tracks does not have the aforementioned limitations \cite{cmsalignment}.
The cosmic ray muon and collision  tracks  are copiously measured  in CMS and
are used to derive the absolute alignment parameters in the CMS
coordinate system down to individual detector modules.
The number and distribution of tracks  define the time interval
and the accuracy  of different alignment parameters in
the track-based alignment.
For example, the alignment of large structures, similar to the
alignment with  LAS, can be performed after a few  hours of
data taking.
In the following we  describe some aspects of the alignment procedures using the LAS data and  particle tracks.

\subsection{Alignment with the laser system}
The LAS  alignment procedure is   based on a few assumptions.
First, we assume that the LAS can measure only
relative displacements of the laser beam profile with respect to some
reference position.
In this study all displacements are derived with respect to the TOB position
because the TOB  holds the alignment tubes and is directly connected to the TST.
The offsets of  laser beam profiles from the reference positions are
thus used to calculate
the variations of alignment parameters, not their absolute values.
The relative  alignment assumes that all tracker components, including
the LAS elements, can move.
\par
The second assumption concerns the  definition of alignable objects and
their parameters.
The  laser beams used for the global alignment allow the
reconstruction of  displacements  of the TOB and TIB
in  $\Delta x$, $\Delta y$, and rotations  $Rx$, $Ry$, $Rz$, while
movement along the $z$-axis is not measured in the barrel due to the
orientation of strips along $z$.
The same laser beams in the TECP and TECM  are used to  reconstruct
$\Delta x$, $\Delta y$ and $Rz$, while other parameters are not
constrained due to the radial orientation of the TEC  strips.
In the LAS alignment procedure, each subdetector is considered as a rigid
body and all  deviations from this model are treated as  systematic
uncertainties.
\par
Further, it is also assumed that the orientation of the laser beams
can vary, for example, due to  temperature variation in the alignment tube
leading to small rotations of the beam splitters.
The direction of each $i$th laser beam   is parameterized  by the two
parameters: the offset $\alpha_i$,  and slope $\beta_i$ in the $\varphi$-$z$
plane. These laser beam parameters  are estimated from the LAS
measurements together with other alignment parameters in one global fit.
Note that the laser beams passing through the  mirrors  or through the
silicon sensors  may have some kinks, but these kinks are
independent of the beam orientation.
The assumption of the straightness of the laser beams   implies that
all of the  optical components of the LAS  have  flat surfaces near the laser beam
spot, such that small displacements  of the LAS components do not
affect the alignment parameters.
\par
\begin{figure*}[thb]
  \centering
   \includegraphics[width=0.44\textwidth]{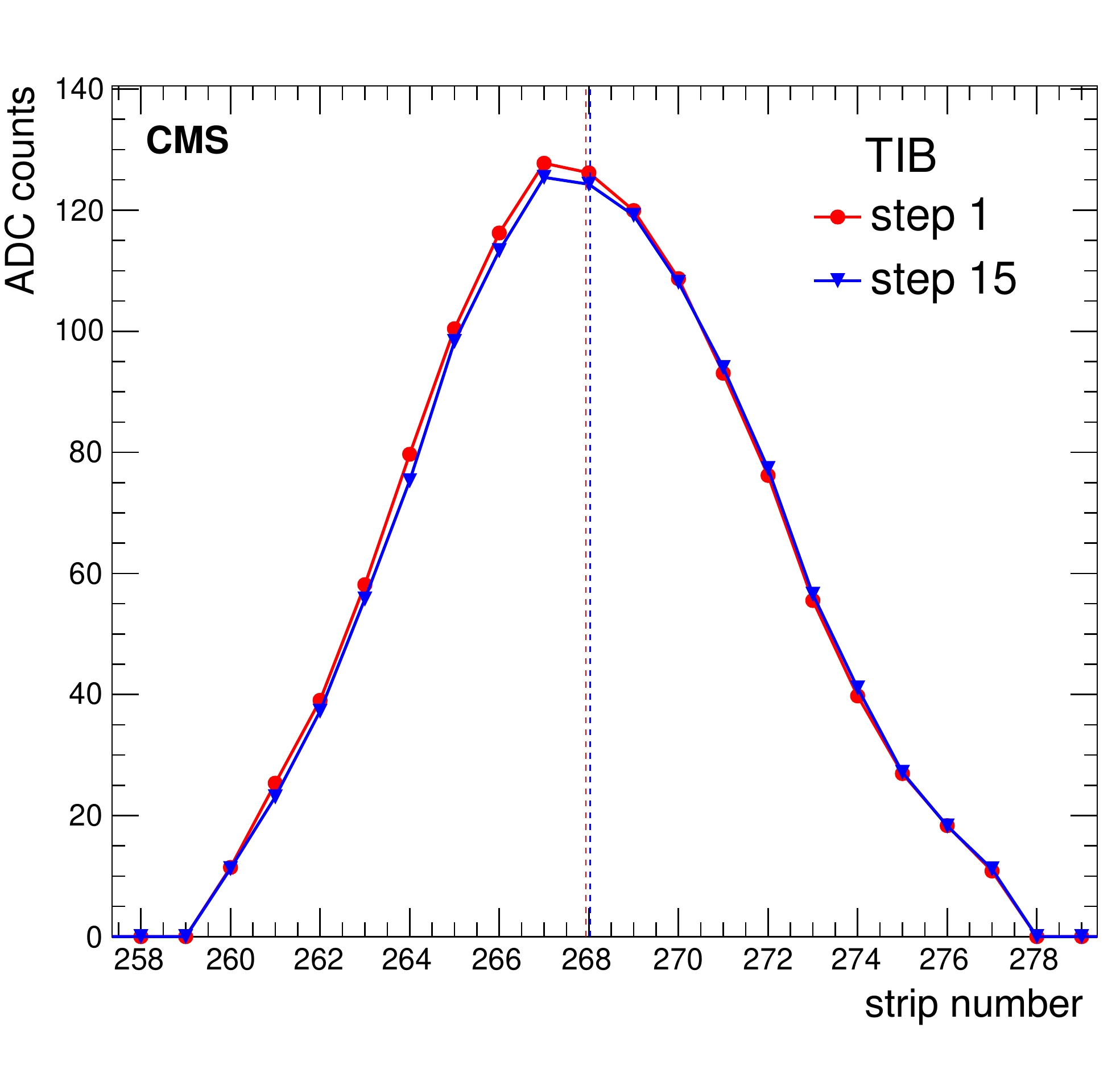}
   \includegraphics[width=0.44\textwidth]{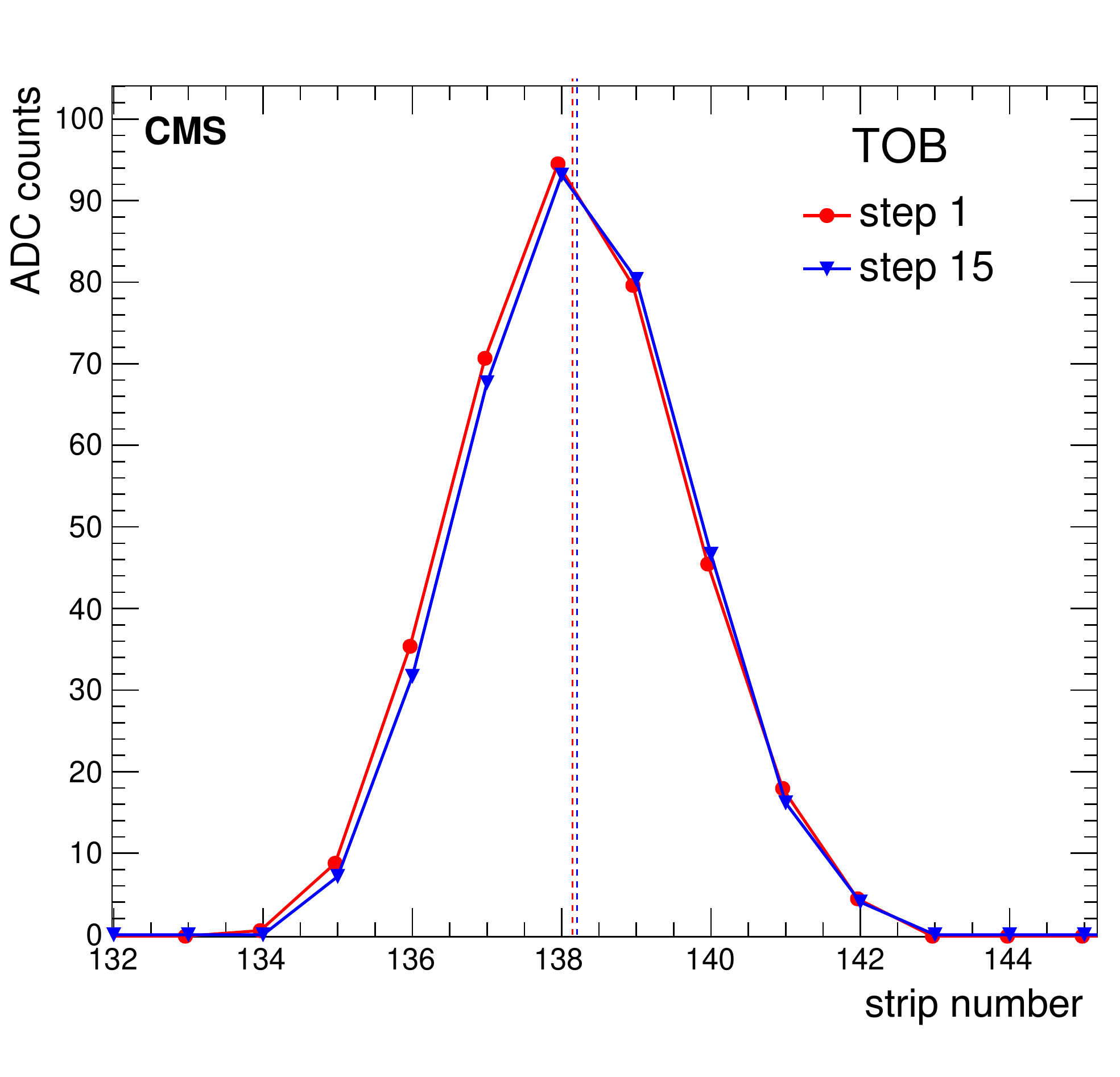}
   \includegraphics[width=0.44\textwidth]{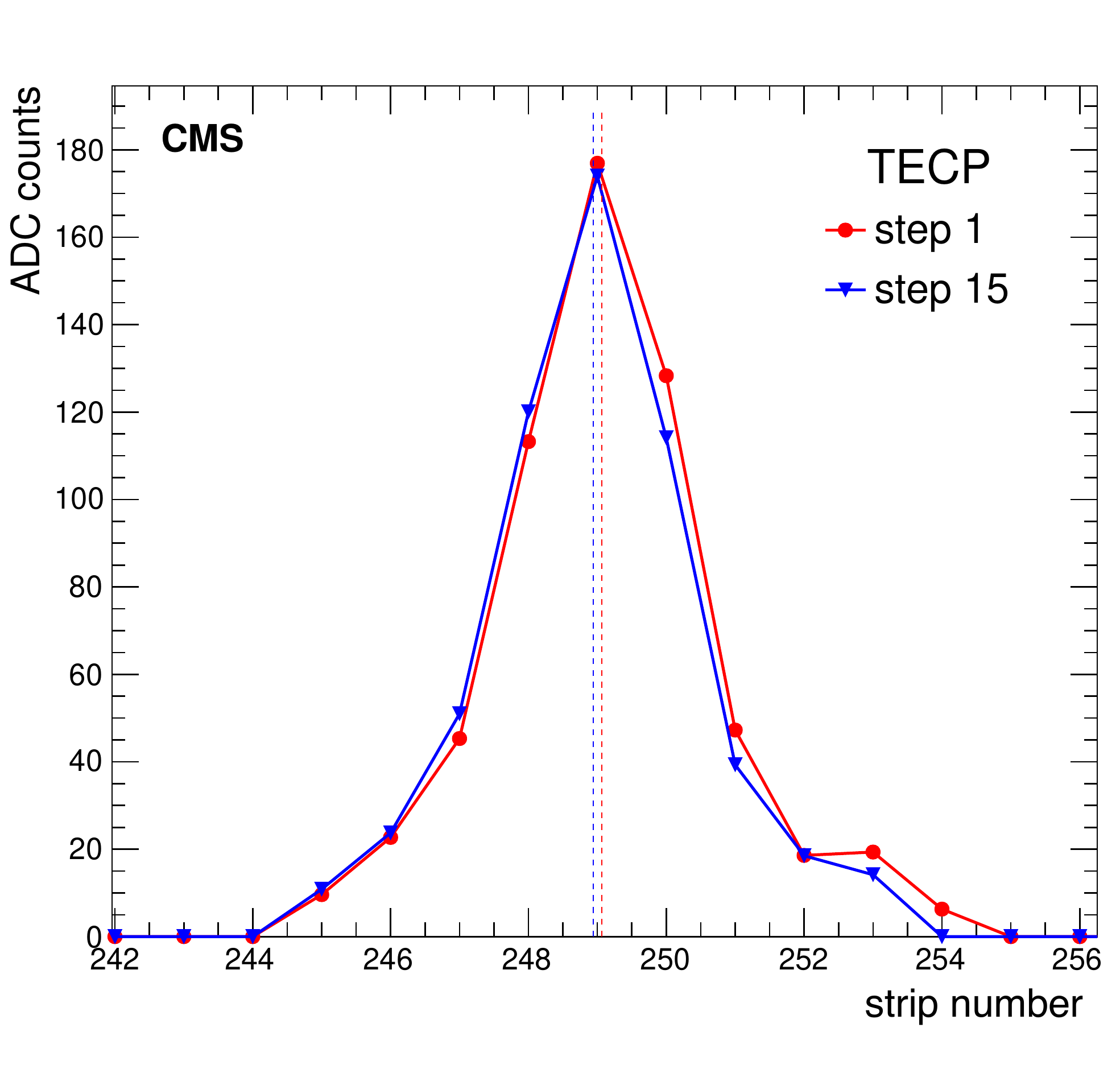}
   \includegraphics[width=0.44\textwidth]{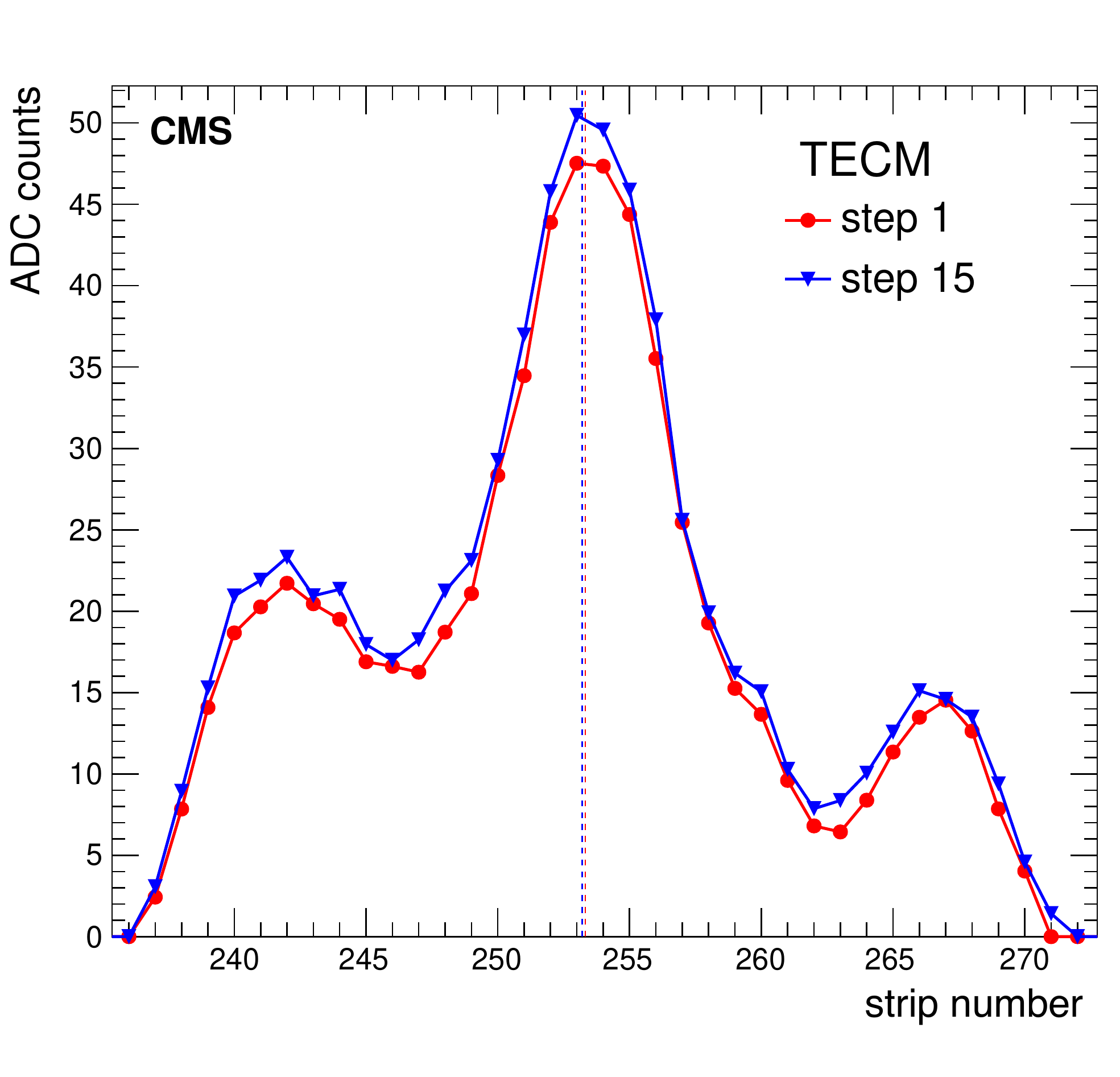}
    \caption{Examples of laser beam profiles (accumulated  amplitude
      in ADC counts for 200 laser shots vs. strip number) in the TIB,
      TOB,  TECP, and  TECM modules, obtained  in two different
      acquisition steps. 
The vertical dashed line shows the
      reconstructed laser spot position. In the TECM, two lower diffractive peaks are visible as mentioned in the text.}
      \label{fig:beamprofiles}
 \end{figure*}

Under the above assumptions the LAS alignment procedure has two main steps: reconstruction of laser beam profiles and evaluation of  alignment parameters.
The laser  profile  is  defined as an accumulated
amplitude in ADC counts  versus strip number within a module after 200 laser shots.
The profile  depends on  the laser intensity, the silicon
strip pitch   and the width of the laser beam spot
 after propagation of laser light through beam splitters, mirrors, and
 silicon sensors (for TECs).
Figure \ref{fig:beamprofiles} shows beam profiles for different
subdetectors obtained in  two acquisition steps. The position of the laser beam spot is obtained from the intersection
of the linear extrapolations of the profile at its  half-maximum.
The strip pitches for
the illuminated TIB, TOB, and TEC modules are the  120, 122, and 128\mum respectively.
For the TOB and TIB  the profiles are Gaussian-like, while for the TEC
detectors, where the light passes through the silicon, the
profiles show a diffraction pattern caused by reflections inside the
silicon sensor.
\par
The second step is the reconstruction of the alignment parameters with
respect to the reference position in each illuminated  module.
The  displacements   of the beam  position  $\psi^{j}$ in  detector
module $j$  with respect  to the reference position $\psi^{j}_\text{ref}$  are given by
$\delta^{j} = \psi^{j}-\psi^{j}_\text{ref}$. These
displacements $\delta^{j}$ are  inputs to the $\chi^2 = \Sigma
{(\delta^{j}-\phi^{j}(\vec{p}))^2}/{\sigma_{j}^2}$, where $\phi(\vec{p})$ are the predicted
displacements depending on the fit parameters $\vec{p}$:  $\alpha_i$, $\beta_i$, $\Delta x$, $\Delta y$, $Rx$, $Ry$, $Rz$.
For  the  $\chi^2$ minimization, the
$\partial \chi^2 / \partial p_{k}$
derivatives
are linearized using the small-angle approximation. The system of
linear equations is solved analytically using matrix inversion
with respect  to  the  parameters $p_{k}$.
The LAS alignment procedure is flexible; some measurements or all
measurements from a specific laser beam
can be excluded from the fit, and the  number of fit parameters  can  be
varied, for example excluding rotations or offsets. These features  were used to check the stability  of the alignment procedure and systematic uncertainties.
The  results from different fit configurations  agree within 10\mum.
\par
The stability of the alignment parameters reconstructed with the LAS  has
been studied during  periods (of a few weeks) of operation at a fixed temperature   where
we expect no real movements of the tracker components.
The stability is defined as one standard deviation of the distribution
of each alignment parameter.
A summary of the LAS alignment parameters
and their stability is presented  in Table ~\ref{tab:laspar}.
The  best stability of  about 1\mum
is obtained for the relative $\Delta x$ and $\Delta y$  displacements  of
the TIB.  For the TECM  profiles the stability  worsens  to 2--3\mum due
to larger distortion of the laser beam after  passing  through many
mirrors.
\begin{table}[htb]
\centering
\topcaption{Stability of alignment parameters using LAS measurements
  obtained during  periods of operation at a  fixed temperature.}
\begin{tabular}{lccccc}
        &  $\Delta x $   &   $\Delta y $       &  $Rx$   &  $Ry$   &  $Rz$  \\
   & [\mum]        & [\mum]         & [$\mu$rad]      & [$\mu$rad]      & [$\mu$rad]      \\ \hline

TIB     &  0.9                     &    0.9                  &     1.7            &    1.6               &  1.1 \\
TECP    & 1.4                      &   1.4                   &      \NA            &     \NA              & 1.7   \\
TECM    & 2.2                      &   2.5                   &    \NA           &     \NA              & 2.9   \\
\end{tabular}
\label{tab:laspar}
\end{table}

\subsection{Alignment with particle tracks}
A  detailed  description of tracker alignment with tracks can be found
in numerous publications, $\eg$ in Ref. \cite{cmsalignment} and references therein.
One of the track-based alignment algorithms used in CMS is
\MILLEPEDE II ~\cite{mpede}. The algorithm simultaneously
reconstructs  the  track parameters $\vec{x}$  for each
event and the  alignment parameters $\vec{p}$ for each alignable object, and involves two steps.
In the first step   the
${\partial f}/{\partial x_i }$
and   ${\partial f}/{\partial p_k }$ derivatives of the track model
$f(\vec{x},\vec{p})$ with respect to the track and alignment
parameters are calculated.
These derivatives are stored in a matrix with the size of
$(n_\text{tracks} n_\text{trackpar} +  n_\text{algnpar})^2$, where $n_\text{tracks}$ is
the number of selected tracks, $n_\text{trackpar}$ is
the number of individual track parameters (four for propagation
without magnetic field and
five for propagation in the field),
$n_\text{algnpar}$ is the  number of alignment parameters.
Then the corresponding system of linear equations is reduced in size using
block matrix algebra  and solved numerically \cite{cmsalignment}.
\par
The phase-space of particle tracks defines the sensitivity of the track-based
alignment procedure to a particular alignment parameter.
Two  types of tracks can be used;
tracks from collisions  that  originate in the detector
centre, and  tracks from cosmic rays  that can cross the detector  away from the
interaction point.
For the 2012 period, about 15$\times$10$^{6}$ collision  tracks and
4$\times$10$^{6}$ cosmic ray tracks  were used for the alignment.
The track samples are split into separate periods in time that are
used to calculate the alignment parameters for all detector modules.
The intervals  should be chosen such that, within each period,  the
operations do not vary significantly, but at the same time should provide
sufficient statistics for  the \MILLEPEDE  procedure. Usually, each interval corresponds to a few months of stable operation.
The reconstruction accuracy of different alignment parameters  in the track-based
alignment depends on the number of selected tracks and on the location of the detector
modules \cite{cmsalignment}.

\section{Tracker  mechanical stability}
The tracker mechanical structures  have a hierarchy, and can be
grouped as follows: subdetectors (TIB, TOB, TECs), substructures (shells,
rods, petals), and individual detector modules. All these components
can potentially move for different reasons and over different time scales.
We distinguish between  short-term variations, which occur over an
interval of a  few hours, and long-term variations, which occur over
a period of a few days or months.
\par
Temperature variation is expected to be the main source of  movement
in the tracker during physics operation.
The thermal expansion  of the CF composite used in the support structures is about 2.6$\times$10$^{-6}$/$^\circ$C.
For the 2.4\unit{m} long TOB this would result in displacements of about
60\mum for $\Delta\mathrm{T} =  10^\circ$C.
Since the mechanical design of the tracker allows for  thermal
expansion,  the temperature-related movements should be elastic, that
is, the positions are restored  when the temperature is restored to
its original value.
However, this process can be disrupted by the uncontrollable   static friction in kinematic joints and
the thermal expansion of  power cables, cooling pipes, etc.  that are
integrated into the structures of the  tracker.
Many thermal cycles of the tracker can thus result in some nonelastic displacements
and non-rigid body deformations.
\par
The release of intrinsic stresses  produced during assembly  is
another source of movement that can happen occasionally  or be initiated by the temperature variations.
Variations of the magnetic field, intervention in the CMS  cavern and mechanical work during technical stops can also
cause the movement of some CMS components and affect the tracker alignment.
These movements, as well as nonelastic movements and deformations, are difficult to simulate in
finite-element method models, thus making experimental  measurements  indispensable for validation of the mechanical design.

\subsection{Long-term stability}
\begin{figure*}[!thb]
  \centering
   \includegraphics[width=0.8\textwidth]{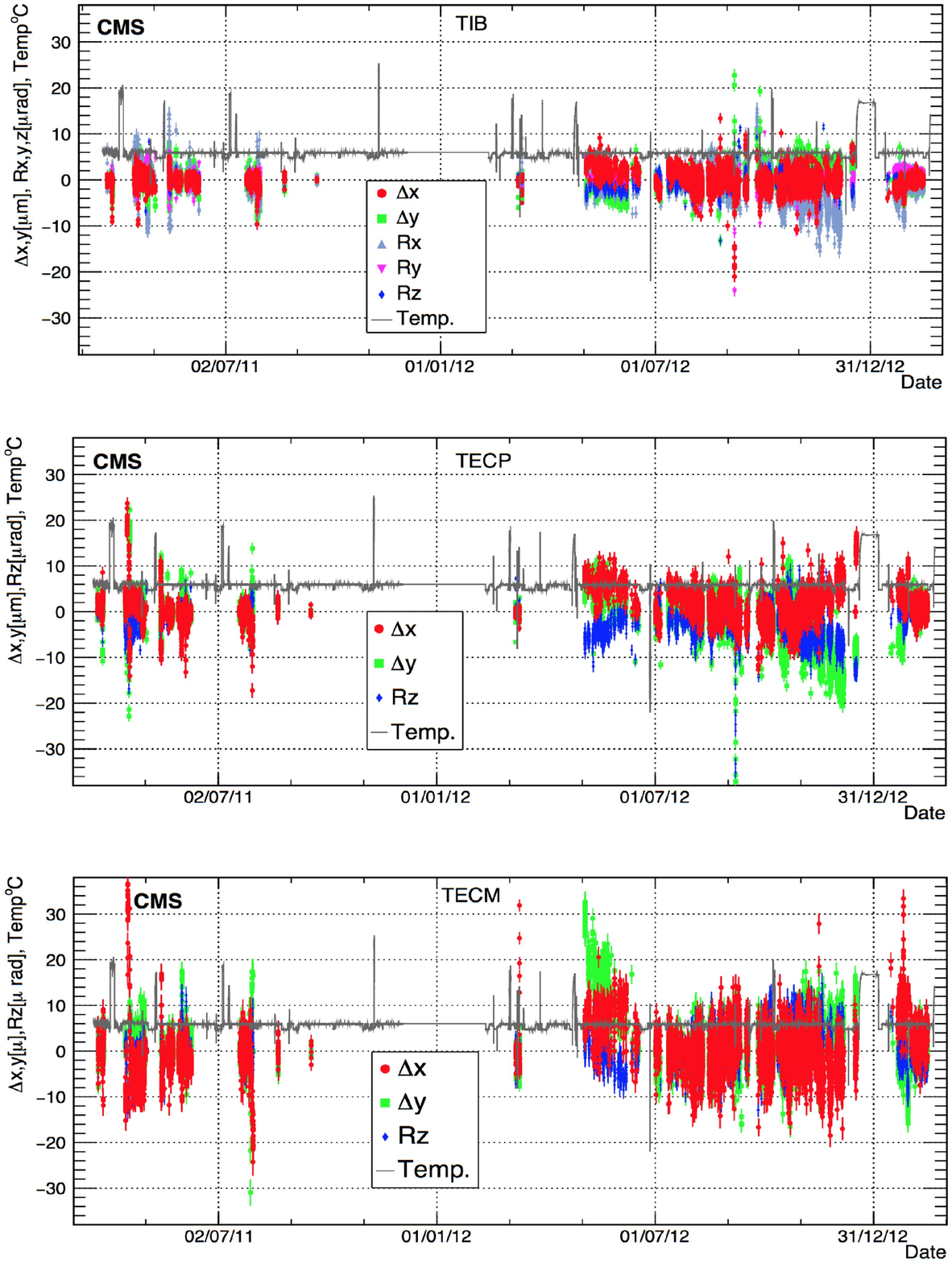}
    \caption{Stability of TIB, TECP, and TECM  alignment parameters during  2011--2013 data taking.
The black line is the temperature of the cooling liquid in the return circuit.
}
      \label{fig:trackerlas}
 \end{figure*}

\begin{figure*}[thb]
  \centering
     \includegraphics[width=0.44\textwidth]{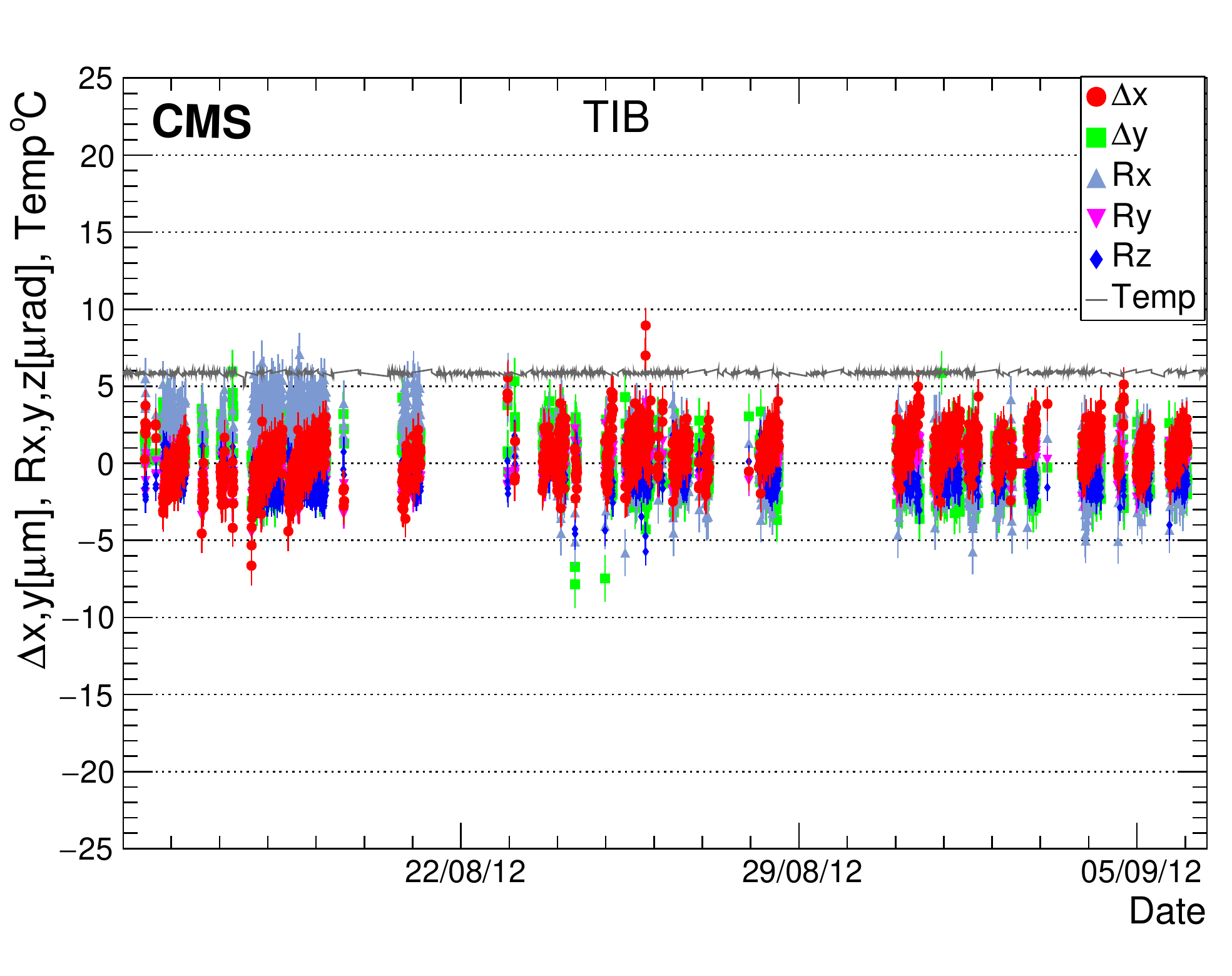}
     \includegraphics[width=0.44\textwidth]{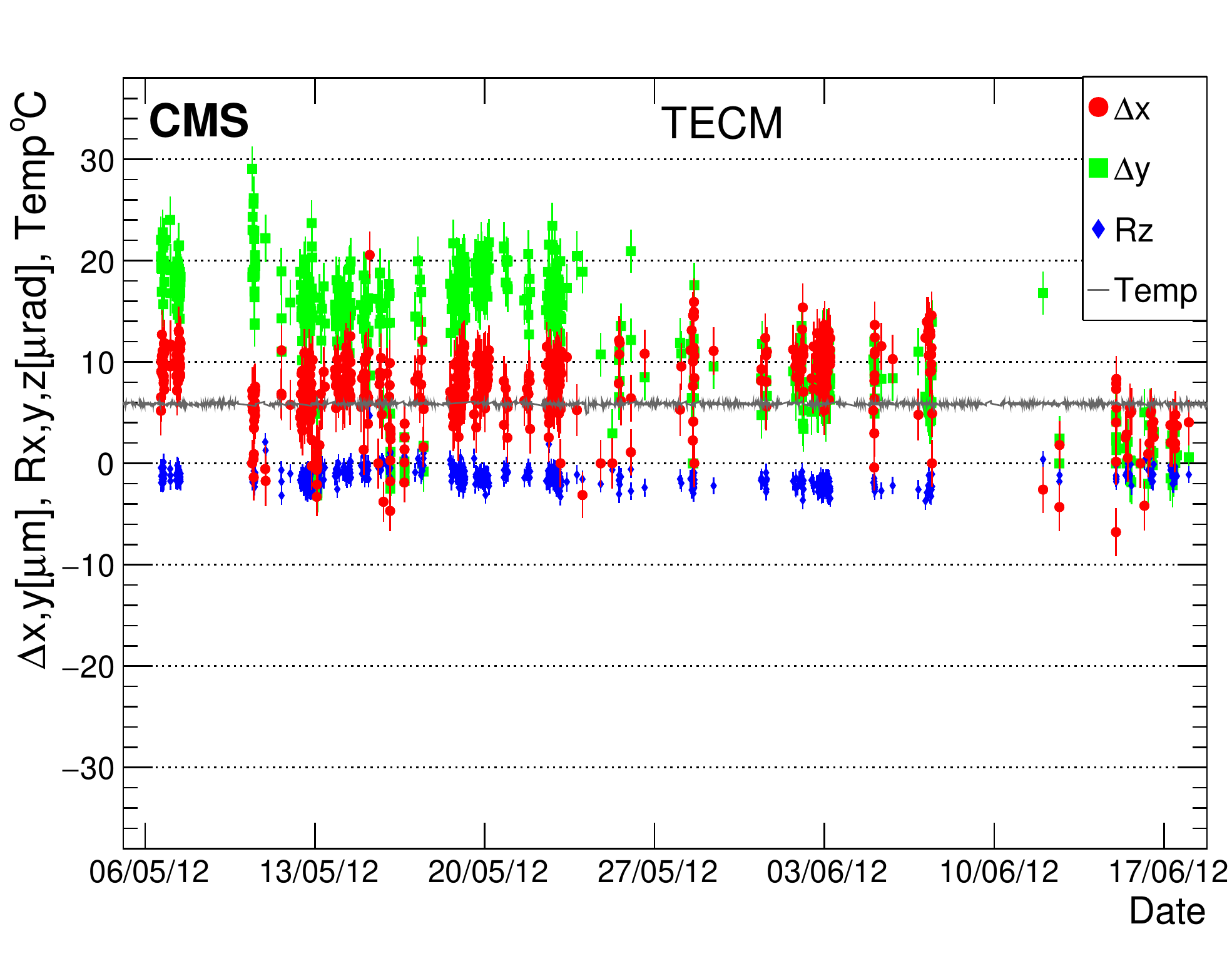}
     \includegraphics[width=0.44\textwidth]{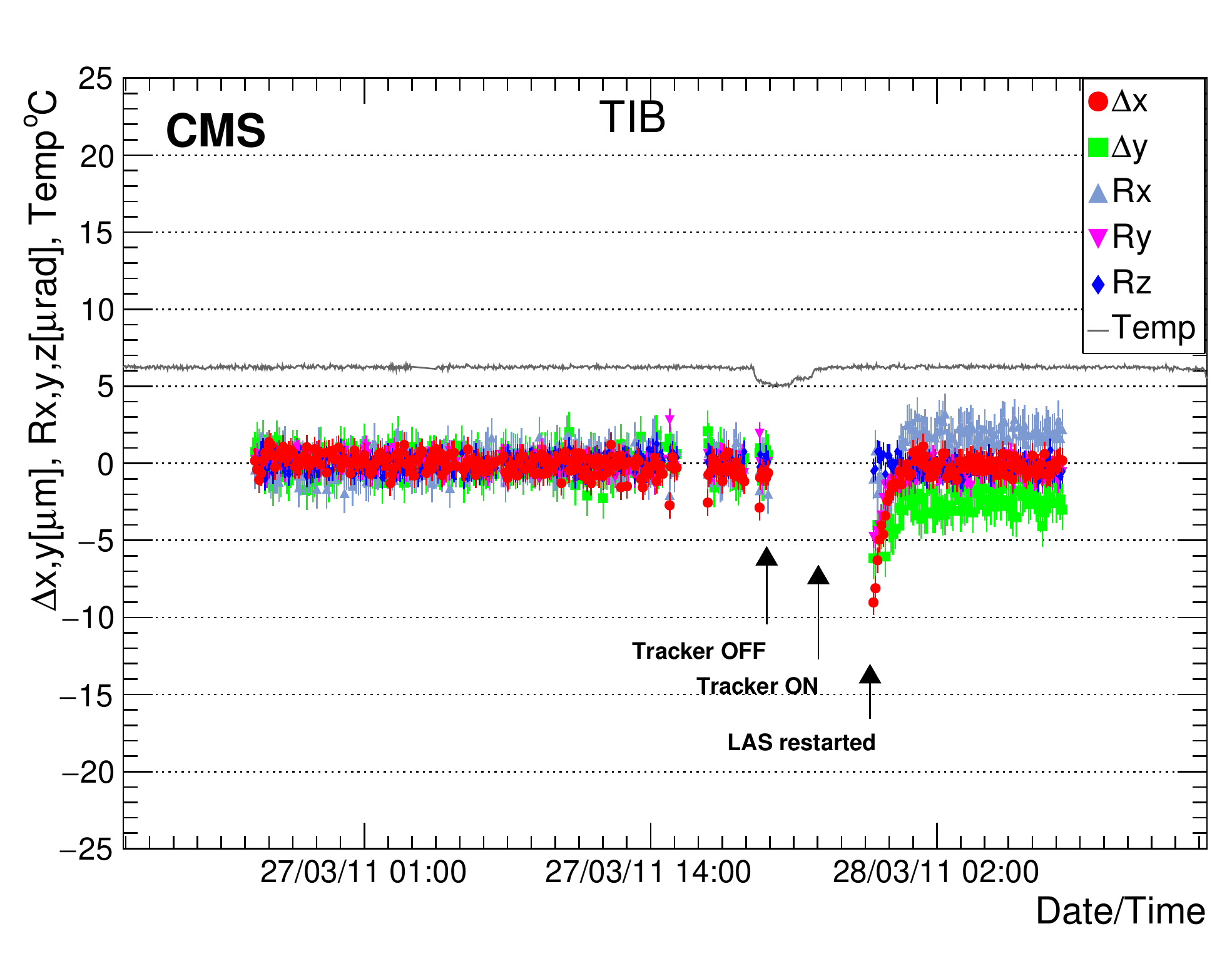}
     \includegraphics[width=0.44\textwidth]{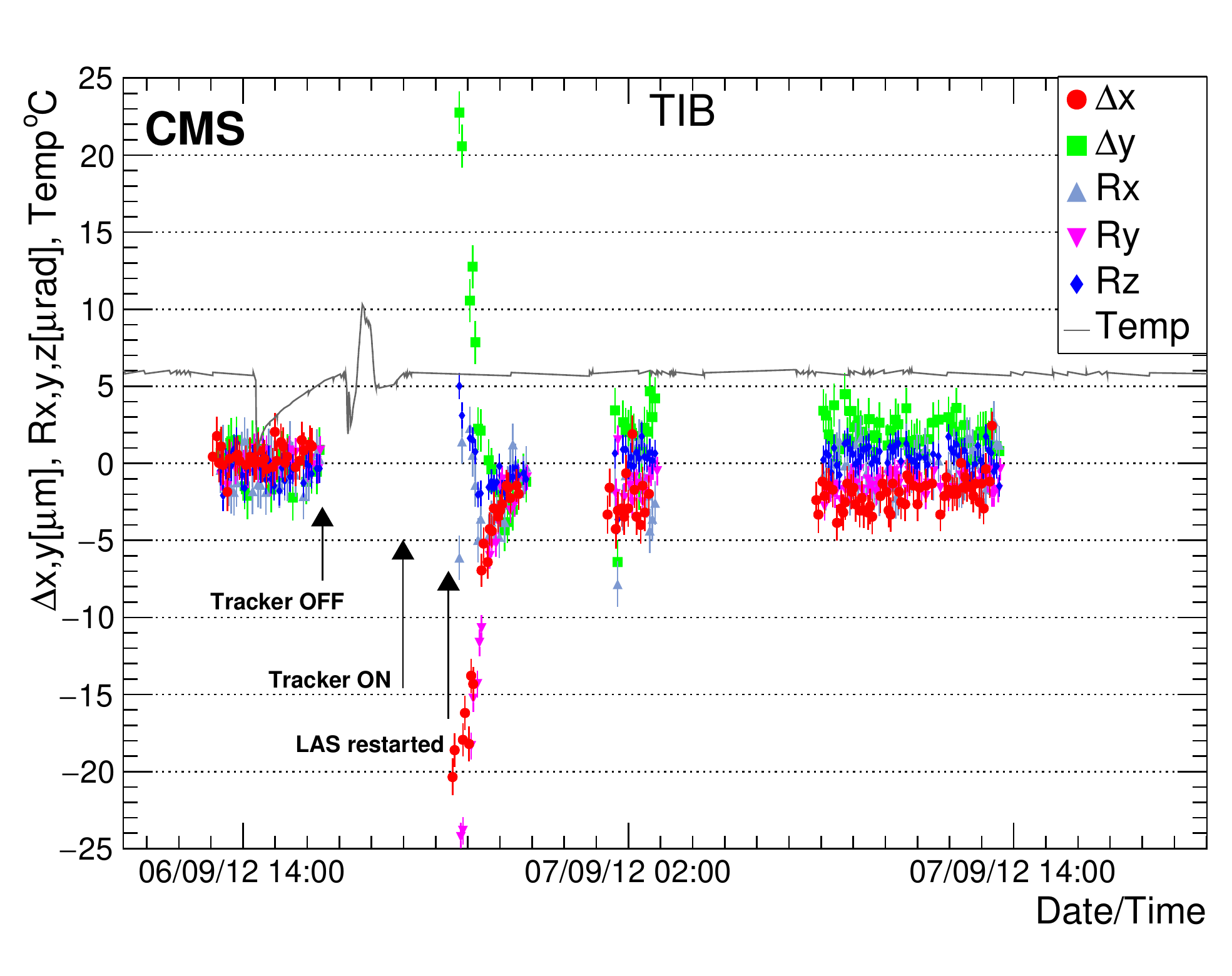}
    \caption{Expanded view of the tracker alignment stability for selected time intervals. Upper left:
      TIB parameters during  weeks of operation at stable temperature.
      Upper right: Variations of TECM parameters at stable temperature.
Bottom left:  Variations of  TIB parameters after a  power trip.
Bottom right: TIB parameters after cooling plant shutdown and recovery.}
      \label{fig:tracklasevents}
 \end{figure*}

\begin{figure*}[!thb]
  \centering
     \includegraphics[width=0.9\textwidth]{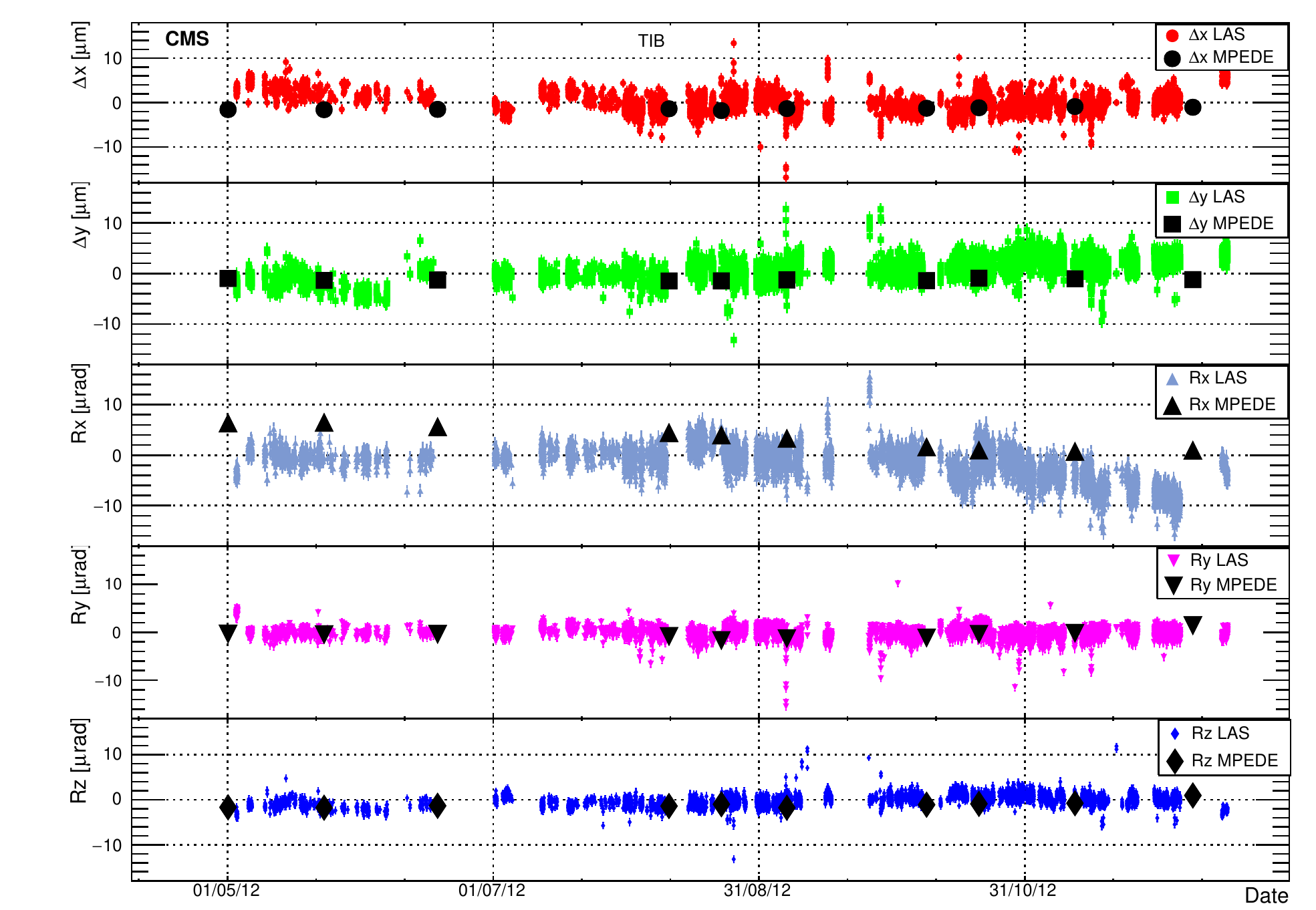}
    \caption{Comparison of the TIB alignment parameters
      reconstructed with LAS data  and calculated with \MILLEPEDE from
      measured particle tracks.}
      \label{fig:lasmpede_tib}
 \end{figure*}

\begin{figure*}[!thb]
  \centering
     \includegraphics[width=0.9\textwidth]{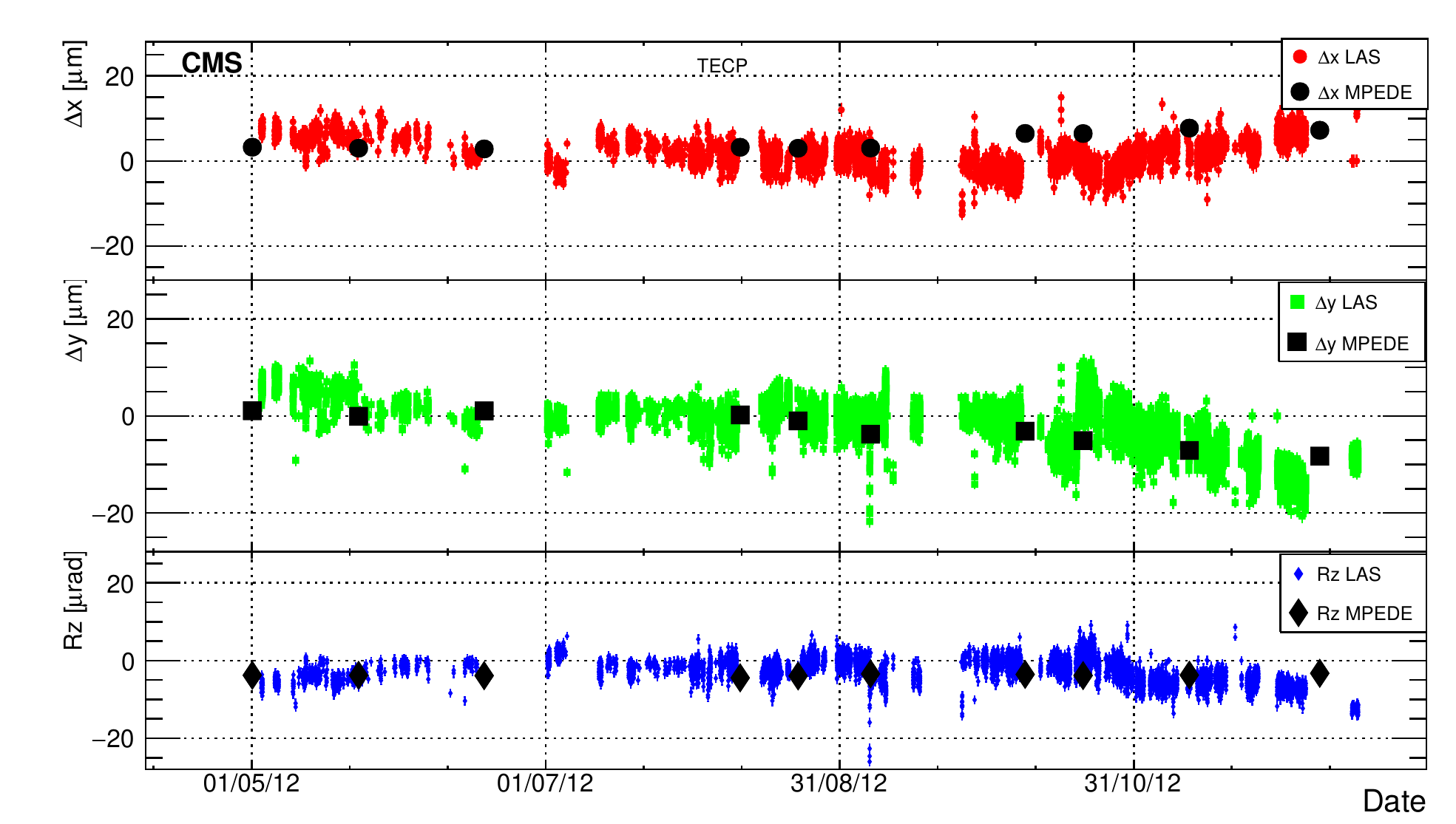}
    \caption{Comparison of TECP alignment parameters measured with
      LAS and calculated with \MILLEPEDE  from
      measured particle tracks.}
      \label{fig:lasmpede_tecp}
 \end{figure*}

\begin{figure*}[!thb]
  \centering
     \includegraphics[width=0.9\textwidth]{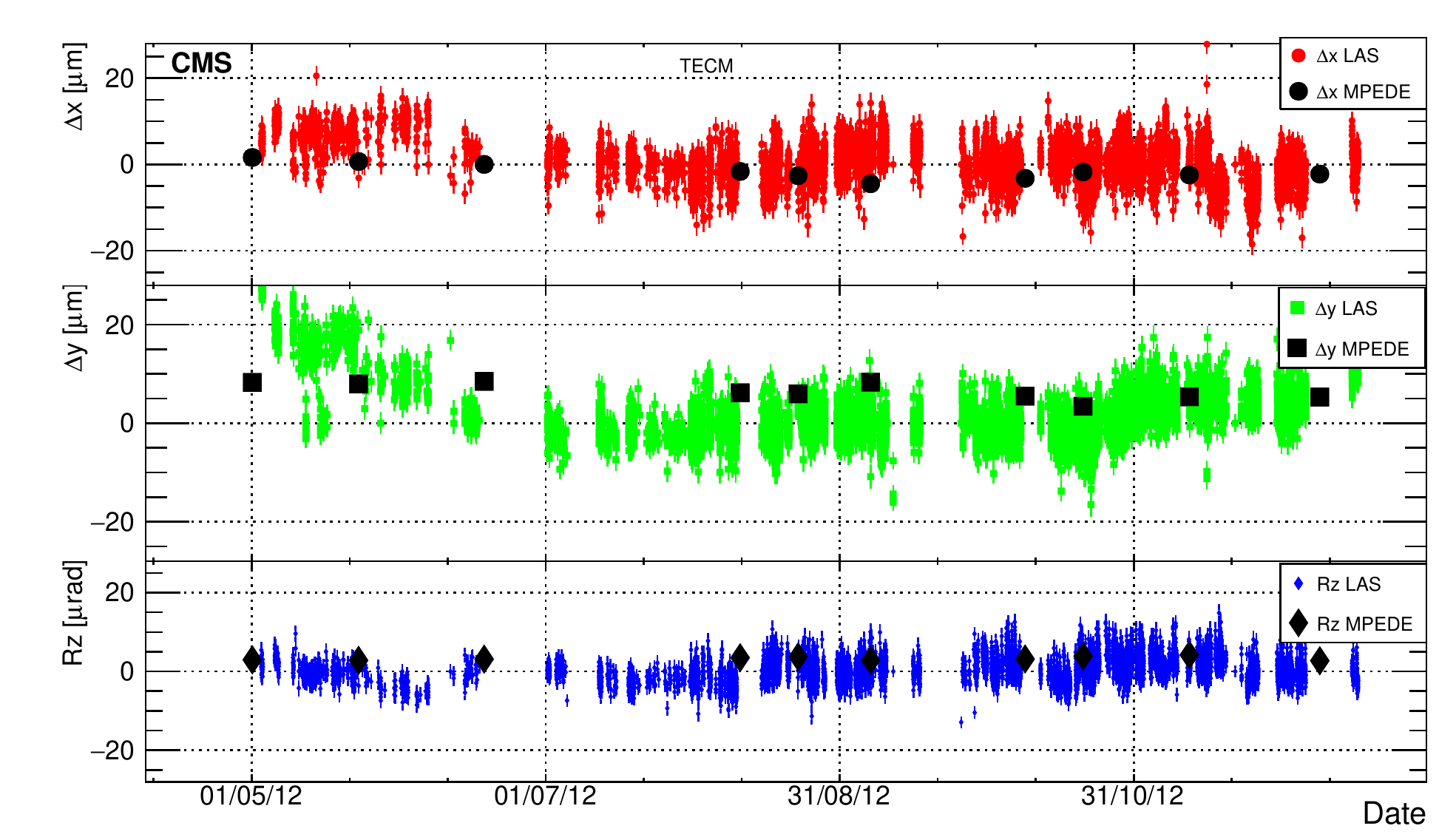}
    \caption{Comparison of TECM alignment parameters measured with
      LAS and calculated with \MILLEPEDE  from
      measured particle tracks.}
      \label{fig:lasmpede_tecm}
 \end{figure*}

The long-term stability of  global alignment parameters reconstructed with
the LAS data in  the years 2011--2013 is shown in
Figure \ref{fig:trackerlas} and in more detail in
Figures \ref{fig:tracklasevents}--\ref{fig:lasmpede_tecm}.
The alignment parameters  of the TIB and TECs are calculated with respect
to the TOB.
Each point in the plots corresponds to  one LAS
acquisition step with an interval of 5 minutes,  and the uncertainties are from the LAS global fit.
Different parameters can overlap in Figure \ref{fig:trackerlas}, but the
range of variations  during the whole period is clearly visible.
The operating  temperature of the cooling plants was set to
$+4^\circ$C throughout the operation  period, resulting in
a temperature of  about $+6^\circ$C in the return pipe, which is
shown as the black line in the figures.
The  positive spikes in the temperature correspond to the occasional power down of the cooling plants, and
the small  negative spikes of about  $2^\circ$C are due to   switching off
 the low-voltage supplies to the detector modules.
The stability of the internal TEC alignment parameters is similar and
is not discussed in this paper.

The whole period of 2011--2013 can be split into different parts.
Periods with no LAS data are due to  either nonoperational  global  CMS DAQ
or  nonoperational tracker.
Loss of data resulting from LAS  problems  was below 1$\%$ and  related to the
occasional powering down of the  LAS  electronics in the service area.
\par
Most of the LAS data were collected during periods of operation  at  stable  temperature inside the tracker volume.
The alignment parameters of the TIB and TEC  are remarkably stable;
all variations in displacements are within
${\pm}10\mum$ for the TIB,  ${\pm}20\mum$ for the TECP, and
${\pm}30\mum$ for the TECM. 
The expanded view of some typical parts shown in  Figure \ref{fig:trackerlas}
can be seen in Figure \ref{fig:tracklasevents},
for example the periods of operation at stable temperature for the TIB
and TECM are presented in the upper  plots.
\par
Stable operation is often interrupted by transient periods when
alignment parameters  change by more than  10\mum (or 10\unit{$\mu$rad})
for TIB and 30\mum (30\unit{$\mu$rad}) for TECs   during  an
interval of  a few hours.
All these periods  are associated  with temperature variations.
The temperature can change rapidly  due to occasional trips of
cooling plants or, more often, due to a power trip affecting some detector
modules. The powering down of the low voltages of the readout hybrids reduces
the temperature locally by about $15^\circ$C.
The actual temperature variations  depend upon how fast the
cooling or voltages are restored, while the  observed variations of the
alignment parameters depend on when the LAS acquisition was
restarted. The bottom left plot in  Figure \ref{fig:tracklasevents}
shows an example of the evolution of the TIB alignment
parameters  after a power trip, affecting the whole tracker.
Power to the tracker was restored and the LAS data acquisition
restarted after 30 minutes, thus the movement during these 30 minutes was
not recorded. The observed evolution  of the alignment  parameters
follows the temperature stabilization in the tracker volume, which
takes about an hour.
Similar effects can be observed during the powerdown of the  cooling plants; in
this case the expected temperature variations and, therefore, the observed
displacements are bigger, as can be  seen in the bottom-right plot in
Figure \ref{fig:tracklasevents}.
The periods with large transient variations of alignment parameters are excluded from physics analysis.
\par
The long periods of  data taking are separated by  a few technical
stops when the whole  CMS detector  is powered down. During this time  the temperature
in the tracker is not  controlled and is close to the ambient
temperature in the detector cavern. At the same time some mechanical
work and intervention to the CMS  detector can take place.
Hence, after each of these technical stops a new reference position is used
in the LAS alignment procedure described above.

\par
The long-term  evolution of the alignment parameters calculated with the LAS
data is compared with the results obtained from the alignment with particle
tracks in Figures \ref{fig:lasmpede_tib}--\ref{fig:lasmpede_tecm}.
The alignment parameters for the 2012 period are calculated for ten
intervals that correspond  to the LAS periods with new reference positions.
The \MILLEPEDE alignment configuration  was similar
to the  configuration for the LAS measurements, that is, the TOB
position was fixed, and the TIB and TEC subdetectors were considered as
rigid bodies that can move
with the same degrees of freedom as used in the LAS.
Since  \MILLEPEDE delivers  absolute alignment parameters based on
measurements in many detector modules, whilst the LAS measures
relative displacements and only for the illuminated modules, some  differences between the parameters
derived with the two different methods are expected.
However the variations  of the parameters in both measurements are
similar and are within 30\mum (or 30\unit{$\mu$rad}),
confirming  the long-term mechanical stability of large structures of the tracker.
The  displacements below 30\mum  can have different origins;
for example they could  be related to deformations of other components of the CMS detector.
Since all   observed  large  transient   variations in the alignment parameters
coincide with  the variations of the temperature in the tracker volume,
a dedicated thermal model  can be used in the future to predict
displacements using solely the temperature measurements.

\begin{figure*}[!thb]
  \centering
\includegraphics[width=0.95\textwidth]{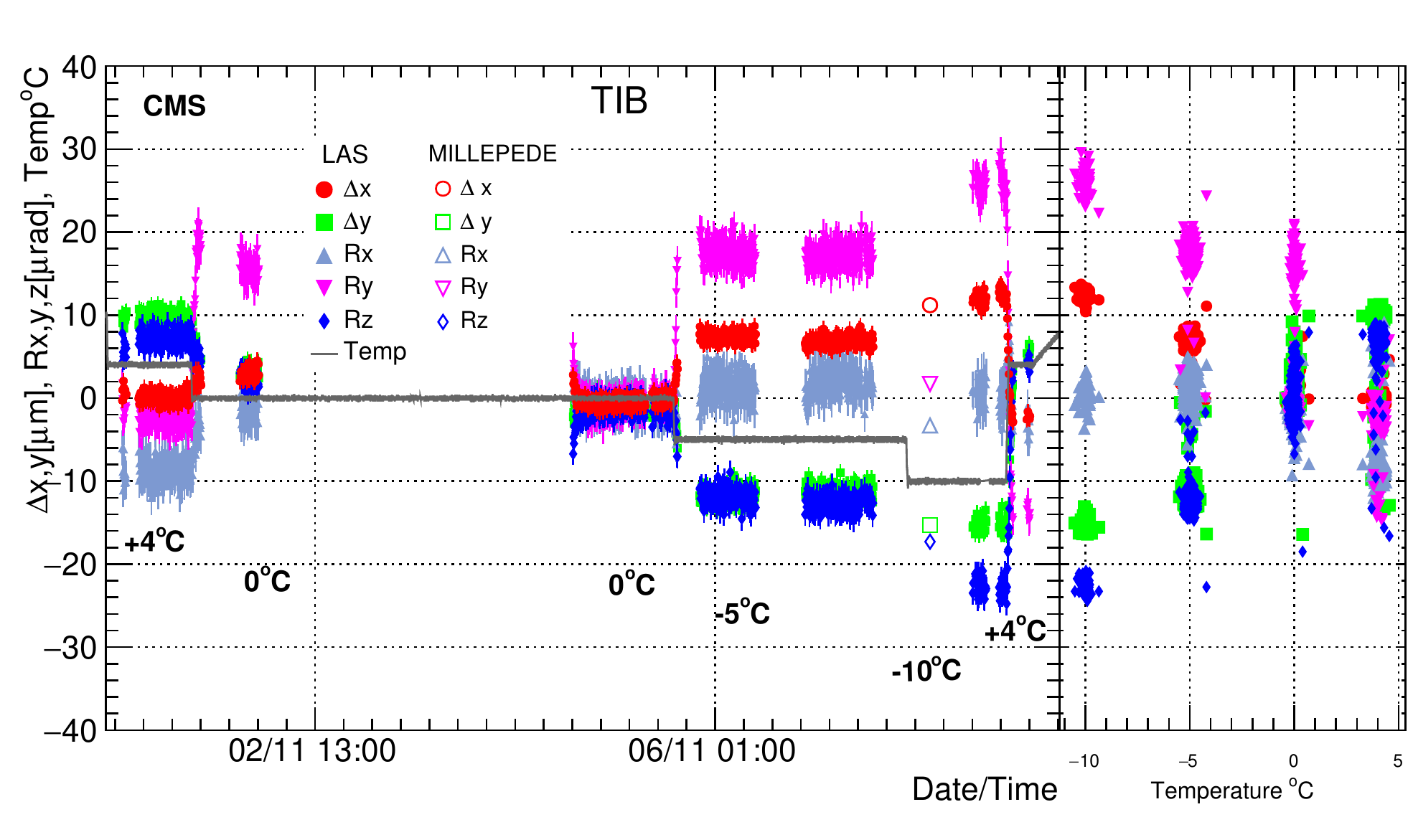}
    \caption{Evolution of the TIB alignment parameters calculated with
      LAS during tracker thermo-cycling  $+4\to 0
      \to-5\to-10\to+4^\circ$C. The \MILLEPEDE  points from cosmic ray
      muons are shown as open markers for the  $0\to -10^\circ$C
      transition.  The right-hand side of the figure groups the data
      at each different temperature, so that the correlation between
      alignment parameter values and operating temperature can be more
      easily distinguished. }
      \label{fig:grinlas}
 \end{figure*}

\subsection{Stability during temperature variations}
\par
Large variations of alignment parameters correlated with the temperature  were
studied during the long shutdown of the   LHC in 2013.
The tracker alignment parameters were reconstructed  when the tracker was cooled down
from a target temperature  of  $+4$ to $-10^\circ$C, and
then warmed  up again.  The evaluation of the TIB alignment
parameters  during temperature transitions from $+4$  to
$-10^\circ$C in steps of  $\Delta \mathrm{T} = 5^\circ$C  is shown in Figure \ref{fig:grinlas}.
The  periods  without data are due to  other CMS commissioning activities that prevented LAS operation.
For all cooldown transitions  the pattern  of  movements is rather
similar; the  parameters change monotonically with
temperature.
When cooled  by $\Delta \mathrm{T} = 5^\circ$C, the TIB $\Delta x$ increases by about
5\mum, $\Delta y$ decreases by 10\mum, and the detector rotates
around the $y$-coordinate  by about 20\unit{$\mu$rad}.  
This corresponds to  temperature-related displacements of 1--2\mum/$^\circ$C.
Some relaxation of
the $Ry$ rotation is observed for the long period at 0$^\circ$C.  Warming up  eliminates most
of the variations immediately, with some remaining residuals of  about
20\mum that were not followed up in this test due to other CMS activities.
\par
The alignment parameters were also calculated with \MILLEPEDE using
tracks from cosmic ray muons, as shown in  Figure \ref{fig:grinlas}.
 About 1.6$\times$10$^3$  and 4$\times$10$^3$ cosmic ray tracks were
recorded for the $0^\circ$C and $-10^\circ$C cooling steps,
respectively. The \MILLEPEDE configuration  was similar to the long-term
stability measurements described in the previous section. The
reference position was taken at $0^\circ$C.
Despite low statistics, both measurements are in reasonable agreement for all
alignment parameters except for the $Ry$ rotation. This rotation
was weakly constrained by cosmic ray  tracks because during the
shutdown only the central part of the muon system was used for the  trigger.

\section{Summary}
The  mechanical stability of the CMS tracker
 was successfully  monitored during the period  2011--2013 using a dedicated
laser alignment system and  particle tracks from collisions and cosmic
ray muons.
During  operation at stable  temperatures,  the
variations of alignment parameters were less than 30\mum.  Larger
changes   were found to be related to temperature variations
caused by the occasional power trip of some modules or of the cooling plant.
These temperature-related displacements of the tracker  subdetectors
are of the order  of 2\mum/$^\circ$C and
are  largely eliminated  when the temperature is restored to its
original value.
\par
The results presented in this study have been crucial for the  CMS tracker
operation in cold conditions. They have established that  major mechanical
displacements do not take place, and have shown the importance of
monitoring the temperature within the detector volume.
The observed behaviour of the tracker components under various
conditions reported here provides guidance for  future
upgrades of the CMS tracking system.

\begin{acknowledgments}
\hyphenation{Bundes-ministerium Forschungs-gemeinschaft Forschungs-zentren Rachada-pisek} We congratulate our colleagues in the CERN accelerator departments for the excellent performance of the LHC and thank the technical and administrative staffs at CERN and at other CMS institutes for their contributions to the success of the CMS effort. In addition, we gratefully acknowledge the computing centres and personnel of the Worldwide LHC Computing Grid for delivering so effectively the computing infrastructure essential to our analyses. Finally, we acknowledge the enduring support for the construction and operation of the LHC and the CMS detector provided by the following funding agencies: the Austrian Federal Ministry of Science, Research and Economy and the Austrian Science Fund; the Belgian Fonds de la Recherche Scientifique, and Fonds voor Wetenschappelijk Onderzoek; the Brazilian Funding Agencies (CNPq, CAPES, FAPERJ, and FAPESP); the Bulgarian Ministry of Education and Science; CERN; the Chinese Academy of Sciences, Ministry of Science and Technology, and National Natural Science Foundation of China; the Colombian Funding Agency (COLCIENCIAS); the Croatian Ministry of Science, Education and Sport, and the Croatian Science Foundation; the Research Promotion Foundation, Cyprus; the Secretariat for Higher Education, Science, Technology and Innovation, Ecuador; the Ministry of Education and Research, Estonian Research Council via IUT23-4 and IUT23-6 and European Regional Development Fund, Estonia; the Academy of Finland, Finnish Ministry of Education and Culture, and Helsinki Institute of Physics; the Institut National de Physique Nucl\'eaire et de Physique des Particules~/~CNRS, and Commissariat \`a l'\'Energie Atomique et aux \'Energies Alternatives~/~CEA, France; the Bundesministerium f\"ur Bildung und Forschung, Deutsche Forschungsgemeinschaft, and Helmholtz-Gemeinschaft Deutscher Forschungszentren, Germany; the General Secretariat for Research and Technology, Greece; the National Scientific Research Foundation, and National Innovation Office, Hungary; the Department of Atomic Energy and the Department of Science and Technology, India; the Institute for Studies in Theoretical Physics and Mathematics, Iran; the Science Foundation, Ireland; the Istituto Nazionale di Fisica Nucleare, Italy; the Ministry of Science, ICT and Future Planning, and National Research Foundation (NRF), Republic of Korea; the Lithuanian Academy of Sciences; the Ministry of Education, and University of Malaya (Malaysia); the Mexican Funding Agencies (BUAP, CINVESTAV, CONACYT, LNS, SEP, and UASLP-FAI); the Ministry of Business, Innovation and Employment, New Zealand; the Pakistan Atomic Energy Commission; the Ministry of Science and Higher Education and the National Science Centre, Poland; the Funda\c{c}\~ao para a Ci\^encia e a Tecnologia, Portugal; JINR, Dubna; the Ministry of Education and Science of the Russian Federation, the Federal Agency of Atomic Energy of the Russian Federation, Russian Academy of Sciences, the Russian Foundation for Basic Research and the Russian Competitiveness Program of NRNU MEPhI (M.H.U.); the Ministry of Education, Science and Technological Development of Serbia; the Secretar\'{\i}a de Estado de Investigaci\'on, Desarrollo e Innovaci\'on and Programa Consolider-Ingenio 2010, Spain; the Swiss Funding Agencies (ETH Board, ETH Zurich, PSI, SNF, UniZH, Canton Zurich, and SER); the Ministry of Science and Technology, Taipei; the Thailand Center of Excellence in Physics, the Institute for the Promotion of Teaching Science and Technology of Thailand, Special Task Force for Activating Research and the National Science and Technology Development Agency of Thailand; the Scientific and Technical Research Council of Turkey, and Turkish Atomic Energy Authority; the National Academy of Sciences of Ukraine, and State Fund for Fundamental Researches, Ukraine; the Science and Technology Facilities Council, UK; the US Department of Energy, and the US National Science Foundation.

Individuals have received support from the Marie-Curie programme and the European Research Council and EPLANET (European Union); the Leventis Foundation; the A. P. Sloan Foundation; the Alexander von Humboldt Foundation; the Belgian Federal Science Policy Office; the Fonds pour la Formation \`a la Recherche dans l'Industrie et dans l'Agriculture (FRIA-Belgium); the Agentschap voor Innovatie door Wetenschap en Technologie (IWT-Belgium); the Ministry of Education, Youth and Sports (MEYS) of the Czech Republic; the Council of Science and Industrial Research, India; the HOMING PLUS programme of the Foundation for Polish Science, cofinanced from European Union, Regional Development Fund, the Mobility Plus programme of the Ministry of Science and Higher Education, the National Science Center (Poland), contracts Harmonia 2014/14/M/ST2/00428, Opus 2014/13/B/ST2/02543, 2014/15/B/ST2/03998, and 2015/19/B/ST2/02861, Sonata-bis 2012/07/E/ST2/01406; the Thalis and Aristeia programmes cofinanced by EU-ESF and the Greek NSRF; the National Priorities Research Program by Qatar National Research Fund; the Programa Clar\'in-COFUND del Principado de Asturias; the Rachadapisek Sompot Fund for Postdoctoral Fellowship, Chulalongkorn University and the Chulalongkorn Academic into Its 2nd Century Project Advancement Project (Thailand); and the Welch Foundation, contract C-1845.
\end{acknowledgments}

\bibliography{auto_generated}

\cleardoublepage \appendix\section{The CMS Collaboration \label{app:collab}}\begin{sloppypar}\hyphenpenalty=5000\widowpenalty=500\clubpenalty=5000\textbf{Yerevan~Physics~Institute,~Yerevan,~Armenia}\\*[0pt]
A.M.~Sirunyan, A.~Tumasyan
\vskip\cmsinstskip
\textbf{Institut~f\"{u}r~Hochenergiephysik,~Wien,~Austria}\\*[0pt]
W.~Adam, E.~Asilar, T.~Bergauer, J.~Brandstetter, E.~Brondolin, M.~Dragicevic, J.~Er\"{o}, M.~Flechl, M.~Friedl, R.~Fr\"{u}hwirth\cmsAuthorMark{1}, V.M.~Ghete, M.~Hoch, C.~Hartl, N.~H\"{o}rmann, J.~Hrubec, M.~Jeitler\cmsAuthorMark{1}, A.~K\"{o}nig, I.~Kr\"{a}tschmer, D.~Liko, T.~Matsushita, I.~Mikulec, D.~Rabady, N.~Rad, B.~Rahbaran, H.~Rohringer, J.~Schieck\cmsAuthorMark{1}, J.~Strauss, W.~Waltenberger, C.-E.~Wulz\cmsAuthorMark{1}
% ...
\vskip\cmsinstskip
% ...
1:~~Also at~Vienna~University~of~Technology, Vienna, Austria\\
% ...

\end{sloppypar}
\end{document}